\documentclass[twocolumn,tighten]{aa}

\usepackage{graphicx}
\usepackage{mathtools}
\usepackage[separate-uncertainty=true,detect-weight,retain-explicit-plus]{siunitx}
\usepackage{hyperref}
\usepackage{cleveref}
\usepackage{caption}
\usepackage{varwidth}
\usepackage{tikz}
\usepackage{enumitem}
\usepackage{microtype}
\usepackage{booktabs}
\usepackage{threeparttable}
\usepackage{enumerate}
\usepackage{multicol}
\usepackage{multirow}
\usepackage{MnSymbol}

\newcommand{\lgalaxies}{\textit{L-Galaxies}}
\newcommand{\emosaics}{\textit{E-MOSAICS}}

\bibpunct{(}{)}{;}{a}{}{,}

\DeclareSIUnit \pc {pc}
\DeclareSIUnit \Msun {M_{\odot}}
\DeclareSIUnit \mag {mag}
\DeclareSIUnit \yr {yr}
\DeclareSIUnit \hubble {\mathit{h}}

\DeclareMathOperator{\erf}{erf}

\creflabelformat{equation}{#2#1#3}

\graphicspath{{./}}

\hypersetup{
    draft=false,
    colorlinks=true,
    allcolors=blue,
    pagebackref=true,
    bookmarks=true,
}

\usetikzlibrary{positioning}
\usetikzlibrary{decorations.pathreplacing}
\usetikzlibrary{bending,calligraphy}

\begin{document} 

\title{Massive Star Clusters in the Semi-Analytical Galaxy Formation Model {\lgalaxies}{\,}2020}
\author{
    \href{https://www.orcid.org/0000-0001-8040-4088}{Nils~Hoyer}\inst{1,2,3,4}\and%
    \href{https://www.orcid.org/0000-0002-6381-2052}{Silvia~Bonoli}\inst{2,5}\and%
    \href{https://www.orcid.org/0000-0001-5679-4215}{Nate~Bastian}\inst{2,5}\and%
    \href{https://www.orcid.org/0009-0007-8173-1161}{Diego~Herrero-Carri{\'{o}}n}\inst{2,6}\and%
    \href{https://www.orcid.org/0000-0002-6922-2598}{Nadine~Neumayer}\inst{3}\and%
    \href{https://www.orcid.org/0000-0002-6143-1491}{David~Izquierdo-Villalba}\inst{7}\and%
    \href{https://www.orcid.org/0000-0002-9074-4833}{Daniele~Spinoso}\inst{8}\and%
    \href{https://www.orcid.org/0000-0001-9320-4958}{Robert~M.~Yates}\inst{9}\and%
    \href{https://www.orcid.org/0000-0002-8889-2167}{Markos~Polkas}\inst{10,2,5}\and%
    \href{https://www.orcid.org/0000-0003-0570-785X}{M.~Celeste~Artale}\inst{11}
}
\institute{
    LIRA, Observatoire de Paris, Université PSL, CNRS, Sorbonne Université, Université Paris Cité, CY Cergy Paris Université, 5 place Jules Janssen, 92195 Meudon, France\\ \email{\href{mailto:nils.hoyer@obspm.fr}{nils.hoyer@obspm.fr}} \and 
    Donostia International Physics Center, Paseo Manuel de Lardizabal 4, E-20118 Donosita-San Sebasti{\'{a}}n, Spain \and 
    Max-Planck-Institut f{\"{u}}r Astronomie, K{\"{o}}nigstuhl 17, D-69117 Heidelberg, Germany \and 
    Universität Heidelberg, Seminarstrasse 2, D-69117 Heidelberg, Germany \and 
    Ikerbasque, Basque Foundation for Science, E-48013 Bilbao, Spain \and 
    University of the Basque Country UPV/EHU, Department of Theoretical Physics, Bilbao, E-48080, Spain \and 
    Institute of Space Sciences (ICE, CSIC), Campus UAB, Carrer de Magrans, E-08193 Barcelona, Spain \and 
    Dipartimento di Fisica, Università degli Studi di Milano Bicocca, piazza della Scienza 3, 20126, Milano, Italy \and 
    Centre for Astrophysics Research, University of Hertfordshire, Hatfield, AL10 9AB, United Kingdom \and 
    School of Physics and Astronomy \& Institute for Gravitational Wave Astronomy, University of Birmingham, Birmingham, B15 2TT, United Kingdom \and 
    Departamento de Ciencias Físicas, Facultad de Ciencias Exactas, Universidad Andres Bello, Fernández Concha 700, Las Condes, Santiago, Chile 
}
\date{Received 28 February, 2025; accepted 23 March, 2026}

\abstract{%
    It is established that there exists a direct link between the formation history of star cluster populations and their host galaxies. However, our lack of understanding of the assembly of star cluster populations impede our ability to use them as tracers of galaxy evolution. In this work we introduce a new variation of the {\lgalaxies} \textit{2020} semi-analytic galaxy formation model that includes the formation of star clusters above $10^{4} \, \textrm{M}_{\odot}$ and probes different physical assumptions that affect their evolution over cosmic time. We use properties of different galaxy components and localised star formation to determine the bound fraction of star formation in disks. After randomly sampling masses from an environmentally-dependent star cluster initial mass function, we assign to each object a half-mass radius, metallicity, and distance from the galaxy centre. We consider up to $2000$ individual star clusters per galaxy and evolve their properties over time taking into account stellar evolution, two-body relaxation, tidal shocks, dynamical friction, and a re-positioning during galaxy mergers. Our simulation successfully reproduces several observational quantities, such as the empirical relationship between the absolute $V$-band magnitude of the brightest young star clusters and the host galaxy star formation rate, the mass function of young star clusters, and mean metallicities of the star cluster distributions versus galaxy masses. The simulation reveals great complexity in the $z=0$ star cluster population resulting from differential destruction channels and origins, including \textit{in-situ} populations in the disk, a major merger-induced heated component in the halo, and accreted star clusters. Model variations point out the importance of the shape of the star cluster initial mass function, the initial distribution of half-mass radii, or the relationship between the sound speed of cold gas and the star formation rate. Our new model provides new avenues to trace individual star clusters and test cluster-related physics within a cosmological set-up in a computationally efficient manner.
}


\maketitle

\section{Introduction}
\label{sec:introduction}

A natural consequence of star formation in cold gas-dense and rotationally unstable regions is the formation of bound stellar structures, ranging from few-body systems to massive star clusters that rival in mass the baryonic components of entire (dwarf) galaxies.
The properties and survival times of these star clusters heavily depend on their changing environment, which makes them excellent tracers of galaxy assembly.
To date, a large body of work has used the present-day star cluster population, such as globular clusters, to constrain the evolutionary history of their current host galaxy, including the Milky Way, M{\,}31, and other nearby systems \citep[e.g.][]{huchra1991a,barmby2000a,perrett2002a,west2004a,brodie2006a,forbes2010b,leaman2013b,huxor2014a,cantiello2015a,veljanoski2015a,myeong2018c,callingham2022a,hammer2023a,hammer2024a,ines-ennis2024a,usher2024a}.
However, despite the clear link between galaxy and star cluster formation that is expressed via, for example, the scaling relation between the number of star clusters and host galaxy mass \citep[e.g.][]{west1995b,blakeslee1999c,peng2008a,spitler2009a,hudson2014a,elbadry2019b,bastian2020c,burkert2020a,zaritsky2022a,le2025a}, precise details about the formation environments and initial properties of $z=0$ star clusters remain elusive \citep[e.g.][]{forbes2018a,valenzuela2025a}.

Constraints on the ages of globular clusters in galaxies, including the Milky Way, suggest that they form at redshifts $z \gtrsim 1$ \citep[e.g.][]{carretta2000a,krauss2003a,lee2003a,kaviraj2005a,strader2005a,correnti2016a}.
Studying the natal environment of clusters at these distances is challenging; however, observations with the \textit{Very Large Telescope}, the \textit{Hubble Space Telescope}, and \textit{James Webb Space Telescope}, combined with strong gravitational lensing, make such observations possible \citep[e.g.][]{vanzella2017c,vanzella2019a,kikuchihara2020a,mowla2022b,forbes2023a,vanzella2023a,claeyssens2023a,mowla2024a}.
Most recently, \citet{adamo2024a} discussed the properties of young and massive star clusters in the ``Cosmic Gems Arc'' at redshift $z \approx \num{10}$ \citep{salmon2018a,bradley2024a}.
Such clusters are compact (half-light radius $\leq 2 \, \textrm{pc}$), massive (stellar mass $\gtrsim 10^{6} \, \textrm{M}_{\odot}$), and could potentially contribute to the $z = 0$ star cluster population.
Irrespective of whether they survive until $z = 0$, they constitute a significant baryonic component of high-$z$ galaxies \citep[$\gtrsim 30 \, \si{\percent}$ for the Cosmic Gems Arc;][]{adamo2024a} and most likely influence the host galaxy's evolution.

Simulations are required to understand the properties of the full star cluster distribution as observations can only trace the brightest / most massive and most unobscured star clusters.
Detailed simulations of star cluster formation and evolution have now become feasible in high-resolution hydrodynamical simulations of galaxy formation, ranging from dwarf galaxies to Milky Way analogues.
These simulations either include star clusters in full cosmological simulations, sometimes including adaptive mesh-refinement or zoom-in techniques \citep[e.g.][]{li2017a,li2018a,li2019a,brown2022a,reina-campos2022b,garcia2023a,calura2025a} or use high spatial- and temporal resolutions but evolve for less than one Gyr \citep[e.g.][]{lahen2020a,hislop2022a,lahen2023a,elmegreen2024c,lahen2024a,lahen2025a,reina-campos2025a}.
Another approach is to add star cluster-related physics to existing simulations.
This approach was realised by e.g.\ the {\emosaics} project \citep[][]{pfeffer2018a,kruijssen2019d} that adds prescriptions for star cluster formation in post-processing to the \textit{EAGLE} simulation \citep{crain2015a,schaye2015a}.
Focussing on {\emosaics}, their simulation can reproduce a number of observables related to the globular clusters mass and metallicity distributions as well as relations with properties of their host galaxy and dark matter halo \citep[see details in][]{hughes2019a,kruijssen2019d,kruijssen2019e,reina-campos2019a,pfeffer2019a,pfeffer2019b,bastian2020c,hughes2020a,kruijssen2020a,reina-campos2020a,keller2020b,horta2021b,trujillo-gomez2021a,dolfi2022a,hughes2022a,reina-campos2022a,reina-campos2023a,pfeffer2024a,newton2025a,pfeffer2025a}.

All of the above mentioned simulations follow in detail the evolution of stars within individual star clusters or of individual baryonic particles, which contain star clusters, but are expensive to run.
In contrast, semi-analytical models sacrifice resolution to lower the computational expense and gain the ability to explore a wide parameter space of various astrophysical mechanisms \citep[e.g.][but see also \citealp{valenzuela2021a} for an empirical model]{white1978a,white1991c,baugh2006a,somerville2008b,somerville2015a}.
This approach has proven to be successful in reproducing observational quantities related to, for example, the co-evolution of galaxies and massive black holes \citep[e.g.][]{kauffmann2000a,croton2006a,croton2006c,de-lucia2007b,monaco2007a,bonoli2009a,bonoli2014a,izquierdo-villalba2020a,gabrielpillai2022a,lagos2024a}, which allows to constrain values of free parameters of the assumed physical models.

For star clusters in a semi-analytical galaxy formation framework specifically, most recently \citet{de-lucia2024a} added prescriptions for their formation to the \textit{GAEA} model \citep{de-lucia2014b,hirschmann2016b}.
The authors were able to reproduce the empirical relationship between the total mass in globular clusters and the parent halo mass \citep[see also e.g.][]{kravtsov2005a,prieto2008d,muratov2010a,li2017a} but utilised simple prescriptions related to mass loss and dynamical friction and relied on global star cluster population statistics.

In this first paper, we introduce an implementation of the formation and evolution of star clusters into a public version of the semi-analytical galaxy formation model ``{\lgalaxies}'' \citep{henriques2020a,yates2021a}.
Our work differs from the above mentioned work by \citet{de-lucia2024a} in that we track the evolution of individual clusters and use different sets of astrophysical prescriptions that make use of the radially-resolved gas and stellar discs in the 2020 version of the code.
This effort enables us to study individual star clusters across different galaxy types, masses, environments, and redshifts, and offers new avenues to study the formation of nuclear star clusters and the co-evolution of black holes with star clusters in future work.

We start in \Cref{sec:model_description} by detailing the governing equations of the model, starting with galaxy components to evaluate the formation efficiency of star clusters, the initial properties of the star clusters, and eventually the evolution of star clusters within the evolution of their host galaxies.
We then evaluate results of our model in \Cref{sec:results} focussing on young massive star clusters in disk-dominated galaxies, and metallicity distributions of \textit{in-situ} and accreted star clusters for different galaxy morphologies, and discuss caveats of our approach.
We conclude in \Cref{sec:conclusions} and present an outlook for future papers in this series.
\Cref{app:sec:model_variations} presents variations of key model parameters.

\section{Model description}
\label{sec:model_description}

Below we outline the basic principles behind our model that we summarise in \Cref{fig:overview}.
Going forward, when mentioning the term {\lgalaxies} we specifically refer to a modified version of the \num{2020} model introduced by \citet{yates2021a}.
This version of the code improves over the default model \citep[][]{henriques2020a} by modifying the prescriptions for metal injection from stellar winds and supernovae into the circum-galactic medium.
\begin{figure*}
    \centering
    \begin{tikzpicture}
        \coordinate (formation) at (0,0);
        \node[font={\Large\bfseries}] at (formation) {Formation};
        \node[below=20pt of formation]{%
            \begin{varwidth}{\linewidth}
                \begin{itemize}[itemsep=1pt, label=$\smalltriangleright$]
                    \item Star formation in galaxy disks (\S~\ref{subsec:the_lgalaxies_semi-analytical_galaxy_formation_model})
                    \item Star cluster formation (\S~\ref{subsec:star_cluster_formation})\\$\rightarrow$ Bound fraction of star formation
                    \begin{itemize}[itemsep=0pt,leftmargin=25pt]
                        \item Cold gas surface density (\S~\ref{subsec:the_lgalaxies_semi-analytical_galaxy_formation_model})
                        \item Epicyclic frequency (\S~\ref{subsubsec:epicyclic_frequency})
                        \item Toomre stability parameter (\S~\ref{subsubsec:toomre_stability_parameter})
                    \end{itemize}
                    \item Star cluster initial mass function (\S~\ref{subsubsec:stellar_mass})\\ Power-law with upper truncation
                    \item Other initial properties:
                    \begin{itemize}[itemsep=0pt]
                        \item Location within annuli (\S~\ref{subsubsec:initial_galactocentric_distances})
                        \item Half-mass radius (\S~\ref{subsubsec:cluster_size})
                        \item Tidal radius (\S~\ref{subsubsec:tidal_radius})
                        \item Metallicity (\S~\ref{subsubsec:metallicity})
                    \end{itemize}
                \end{itemize}
            \end{varwidth}
        };

          \coordinate (evolution) at (10,0);
          \node[font={\Large\bfseries}] at (evolution) {Evolution};
          \node[below=20pt of evolution]{%
            \begin{varwidth}{\linewidth}
                \begin{itemize}[itemsep=1pt, label=$\smalltriangleright$]
                    \item Mass loss (\S~\ref{subsubsec:mass-loss})
                    \begin{itemize}[itemsep=0pt]
                        \item Stellar evolution
                        \item Two-body relaxation
                        \item Tidal shocks
                    \end{itemize}
                    \item Radial expansion (\S~\ref{subsubsec:radial_expansion})
                    \begin{itemize}[itemsep=0pt]
                        \item Two-body relaxation
                        \item Tidal shocks
                    \end{itemize}
                    \item Dynamical friction (\S~\ref{subsubsec:galactocentric_migration})
                    \item Re-distribution during galaxy mergers (\S~\ref{subsubsec:re-distribution_during_galaxy_mergers})
                    \begin{itemize}[itemsep=0pt]
                        \item Minor mergers: accreted clusters move to halo
                        \item Major mergers: all clusters move to halo
                    \end{itemize}
                \end{itemize}
            \end{varwidth}
        };

        \draw[decoration={calligraphic brace,mirror,raise=0pt,amplitude=5pt},decorate,thick] (-100pt,-20pt) -- node[left=6pt] {\rotatebox{90}{Evaluated for each annulus}} (-100pt,-120pt);
    \end{tikzpicture}
    \caption{%
        Relevant prescriptions for the assembly and evolution of star clusters implemented into a modified version \citep[][]{yates2021a} of {\lgalaxies} \num{2020} \citep[][]{henriques2020a}.
    }
    \label{fig:overview}
\end{figure*}

\subsection{The {\lgalaxies} semi-analytical galaxy formation model}
\label{subsec:the_lgalaxies_semi-analytical_galaxy_formation_model}

The {\lgalaxies} model combines merger trees from dark matter-only $N$-body simulations with a set of partial differential equations for the evolution of baryonic components.
It has been developed to primarily run on the Millennium \citep{springel2005c} and Millennium-II \citep{boylan-kolchin2009a} simulations with box sizes / dark matter particle masses of $480.3 \, h^{-1}\textrm{Mpc}$ / $9.61 \times 10^{8} \, h^{-1} \textrm{M}_{\odot}$ and $96.1 \, h^{-1} \textrm{Mpc}$ / $7.69 \times 10^{6} \, h^{-1} \textrm{M}_{\odot}$, respectively.
Dark matter (sub-)halos are identified using a ``Friends-of-Friends'' \citep{davis1985a} and the ``subfind'' algorithm \citep{springel2001c,dolag2009d} and are used as input to {\lgalaxies}.
As discussed in the works related to the last few major releases \citep{guo2011b,henriques2015a,henriques2020a}, the model can reproduce many observables of baryonic components, such as the redshift-dependent galaxy mass function, passive galaxy fraction, and the cosmic density of the star formation rate (SFR).

Many extensions to \textit{L-Galaxies} have been developed, including those that focus on the gas \citep{vijayan2019a,ayromlou2021b,yates2021a,parente2023a,zhong2023a,zhong2023b,parente2024a,yates2024a}, stars \citep{bluck2016a,wang2018c,irodotou2019a,izquierdo-villalba2019a,murphy2022a,wang2024e}, massive black holes \citep{bonoli2009a, bonoli2014a, izquierdo-villalba2020a,izquierdo-villalba2022a,spinoso2023a,polkas2024a}, and other components \citep{barrera2023a,vani2025a}.
This shows that {\lgalaxies} is a versatile utility to explore the assembly of galaxies and has the potential to investigate the formation of star clusters as well.

One of the key features of the $2020$ version of {\lgalaxies} that was adapted from \citet{fu2013a} is the introduction of concentric annuli.
By default, the model features twelve such annuli that are logarithmically-spaced and act as the resolution limit for the cold gas and stars within a galaxy's disk and stars within a galaxy's bulge.
The annuli's outer radii have values of
\begin{equation}
    w_{j} \,/\, [h^{-1} \textrm{kpc}] = 0.01 \times 2^{j} \quad \mathrm{with} \; j \in [1 ,\, 12] \;,
    \label{equ:ring_distances}
\end{equation}
resulting in $w_{1} \approx 29.7 \, \textrm{pc}$ and $w_{12} \approx 60.8 \, \textrm{kpc}$.

One of the affected properties of the separation into annuli is the star formation prescription.
{\lgalaxies} assumes that the molecular gas in each annulus collapses on a dynamical time-scale $\tau_{\mathrm{dyn}}$ and is transformed to stars with an efficiency $\epsilon_{\mathrm{H}_{2}}$ \citep[e.g.][]{fu2012a}.
Thus, in terms of surface mass density, the SFR for ring $j$ is
\begin{equation}
    \Sigma_{\mathrm{SFR} ,\, j} = \epsilon_{\mathrm{H}_{2}} \, \tau_{\mathrm{dyn}}^{-1} \, \Sigma_{\mathrm{H}_{2} ,\, j} \;.
    \label{equ:sfr_density}
\end{equation}
The model assumes for the dynamical time
\begin{equation}
    \tau_{\mathrm{dyn}} = R_{\mathrm{g}} \,/\, v_{\mathrm{max}} \;,
    \label{equ:dynamical_time}
\end{equation}
relating the disk scale-length of the cold gas $R_{\mathrm{g}}$ and the maximum value of the rotation velocity of the dark matter halo $v_{\mathrm{max}}$.
The amount of molecular gas itself is derived from the available cold gas in each ring and, in turn, is related to the gas' metallicity and clumping of gas clouds.
Extensive discussions and recipes for computing the molecular mass (per annuli) are presented in \citet{krumholz2009b,fu2010a,mckee2010a,fu2013a,henriques2020a}.

In this work we adopt the $2014$ cosmology of \textit{Planck} \citep{planck2014a} with $\Omega_{\Lambda, \, 0} = 0.685$, $\Omega_{\mathrm{M}, \, 0} = 0.315$ (with $\Omega_{\mathrm{B}, \, 0} = 0.0487$), $\sigma_{8} = 0.826$, $n_{\mathrm{s}} = 0.96$, and $h = 0.673$ throughout the paper.
\textit{L-Galaxies} itself utilises dark matter-only simulations that are re-scaled \citep{angulo2010b,angulo2015a} to these values.

\subsection{Gravitational potential}
\label{subsec:gravitational_potential}

The computation of the fraction of star formation that is bound in star clusters (\Cref{subsubsec:bound_fraction_of_star_formation}) requires knowledge of the underlying gravitational potential.
We consider the contribution of all available galaxy components that {\lgalaxies} tracks over time, namely a central massive black hole, dark matter, a bulge, gaseous and stellar disks, and gaseous and stellar halos.

\paragraph{\textit{Massive black hole}}
In case a massive black hole occupies a galaxy centre, we introduce a gravitational potential of a point mass, i.e.\
\begin{equation}
    \Phi_{\mathrm{BH}}(w) = -\frac{\mathrm{G} \, M_{\mathrm{BH}}}{w} \;,
    \label{equ:gravitational-potential-black-hole}
\end{equation}
where $G$ is the gravitational constant, $M_{\mathrm{BH}}$ the black hole mass, and $w$ the galactocentric distance.
We assume a constant alignment between the dark matter and galaxy centres and that the massive black hole does not `wander' within the galaxy \citep[see][for a discussion on wandering black holes in {\lgalaxies}]{izquierdo-villalba2020a,untzaga2024a}.

\paragraph{\textit{Dark matter halo}}
We select the classical NFW profile \citep{navarro1996a} to describe the distribution of dark matter.
The gravitational potential is given by
\begin{equation}
    \Phi_{\mathrm{DM}}(w) = -\frac{\mathrm{G} \, M_{\mathrm{vir}}}{w} \, \frac{\ln(\num{1} + c_{\mathrm{vir}} \, w / R_{\mathrm{vir}})}{\ln(\num{1} + c_{\mathrm{vir}}) - c_{\mathrm{vir}} / (\num{1} + c_{\mathrm{vir}})} \;,
    \label{equ:gravitational-potential-dark-matter}
\end{equation}
where we introduced the virial mass $M_{\mathrm{vir}}$ and radius $R_{\mathrm{vir}}$, respectively.
For the concentration parameter $c_{\mathrm{vir}}$ which relates the profile's scale-radius to $R_{\mathrm{vir}}$, we assume a relation from \citet{dutton2014b} that connects it to $M_{\mathrm{vir}}$ via
\begin{equation}
    \log_{10} \, c_{\mathrm{vir}} = \alpha(z) + \beta(z) \, \log_{10} ( M_{\mathrm{vir}} \,/\, 10^{12} \, h^{-1} \textrm{M}_{\odot} ) \;.
    \label{equ:gravitational-potential-dark-matter-c}
\end{equation}
The redshift-dependent coefficients $\alpha(z)$ and $\beta(z)$ take values according to Table~\num{3} of \citet{dutton2014b} with $\alpha(\num{0}) = \num{1.025}$ and $\beta(\num{0}) = \num{-0.097}$.

\paragraph{\textit{Galactic bulge}}
We model a galaxy's bulge with a \citet{jaffe1983a} profile of the form
\begin{equation}
    \Phi_{\mathrm{B}}(w) = -\frac{\mathrm{G} \, M_{\mathrm{B}}}{w_{\mathrm{B}}} \ln (\num{1} + w_{\mathrm{B}} / w) \;,
    \label{equ:gravitational-potential-bulge}
\end{equation}
with a scale-length $w_{\mathrm{B}}$ that encloses half of the bulge's mass $M_{\mathrm{B}}$.

\paragraph{\textit{Gaseous and stellar disks}}
We follow the default assumption in {\lgalaxies} that both the gaseous and stellar disks are well described by two-dimensional exponential density profiles.
The mid-plane gravitational potential of the two disks is expressed with modified Bessel functions of the first and second kind, $I_{\nu}$ and $K_{\nu}$, respectively \citep[e.g.][]{watson1944a,kuijken1989a}.
The gravitational potential for disk $i$ reads
\begin{equation}
    \Phi_{\mathrm{D} ,\, i}(w) = -\pi \mathrm{G} \, \Sigma_{\mathrm{D} ,\, i} \, w \, \big[ I_{0}(y_{i}) K_{1}(y_{i}) - I_{1}(y_{i}) K_{0}(y_{i}) \big] \; ,
    \label{equ:gravitational-potential-disk}
\end{equation}
where $y_{i} = w / (2 \times w_{\mathrm{D} ,\, i})$, $w_{\mathrm{D} ,\, i}$ a characteristic scale-length of disk $i$ (gaseous or stellar), and $\Sigma_{\mathrm{D} ,\, i}$ as the central surface mass density \citep{freeman1970a,binney2008a}.

\paragraph{\textit{Hot gas and stellar haloes}}
We assume that a galaxy's halo contains both hot gas that is unable to form stars and a stellar medium that originates entirely from stripped satellite galaxies.
For simplicity, we assume that an isothermal profile can describe well both the gaseous and stellar halo, i.e.\
\begin{equation}
    \Phi_{\mathrm{H} ,\, i}(w) \propto -\frac{\mathrm{G} \, M_{\mathrm{H} ,\, i}}{R_{\mathrm{H} ,\, i}} \ln (w) \;.
    \label{equ:gravitational-potential-halo}
\end{equation}
Furthermore, we assume that all mass is enclosed within the virial radius such that $R_{\mathrm{H} ,\, i} = R_{\mathrm{vir}}$.\footnote{\citet{yates2017a,yates2024a} discuss and utilise more physically motivated profiles within {\lgalaxies}.}

\subsection{Star cluster formation}
\label{subsec:star_cluster_formation}

The total mass of newly formed stars that are bound in star clusters equals the SFR multiplied by the simulation time step and the cluster formation efficiency \citep[][]{bastian2008b}.
In principle, the latter must be computed considering local conditions of the gas phase, mergers from substructures, and accretion of gas during cluster formation \citep[see e.g.][]{karam2022a,karam2023a}.
This step should ideally be computed at sub-parsec resolution \citep[][]{renaud2020a} with a star-by-star treatment of feedback \cite[][]{calura2025a} but this is not feasible with our approach.

Here we follow the prescription outlined by \citet{kruijssen2012d} to estimate the cluster formation efficiency.
The model relies on three quantities: (1) the cold gas surface mass density $\Sigma_{\mathrm{g}}$, which we compute as an annuli's molecular mass divided by it's area, (2) the epicyclic frequency $\kappa$, and (3) the Toomre disk stability parameter $Q$.
We compute these values for the logarithmic mean galactocentric distance of each annuli $j$ and assume that it is applicable to a wide range of environments and redshifts.\footnote{Using the same cluster formation efficiency calculations from \citet{kruijssen2012d}, \citet{pfeffer2025a} recently showed that the {\emosaics} simulation can well reproduce observed properties of star clusters at high-$z$.}
Furthermore, we assume that all star clusters form in galaxy disks, thus, neglecting cluster formation during galaxy mergers or outside of dark matter halos at high-$z$ \citep[][]{lake2021a,lake2023b}.

\subsubsection{Epicyclic frequency}
\label{subsubsec:epicyclic_frequency}

Based on the combined gravitational potential of all previously introduced galaxy components, $\Phi_{\mathrm{tot}}(w) = \sum_{c} \Phi_{c}(w)$, we can easily determine the epicyclic frequency for circular orbits.
For annulus $j$ and log-mean distance $\langle w_{j} \rangle$,
\begin{equation}
    \kappa_{j} = \sqrt{ \frac{3}{\langle w_{j} \rangle} \frac{\partial \Phi_{\mathrm{tot}}}{\partial w} \bigg|_{\langle w_{j} \rangle} + \frac{\partial^{2} \Phi_{\mathrm{tot}}}{\partial w^{2}} \bigg|_{\langle w_{j} \rangle} } \;.
    \label{equ:epicyclic_frequency}
\end{equation}
Note that, when evaluating this property for galaxy disks, we assume co-rotation of the gaseous and stellar disks resulting in only a single value of these quantities per annuli.

\subsubsection{Toomre stability parameter}
\label{subsubsec:toomre_stability_parameter}

The Toomre stability criterion \citep{safronov1960a,toomre1964a} evaluates whether a disk is stable against collapse considering gravity, pressure, and shear.
For the gaseous and stellar disks we determine
\begin{subequations}
    \begin{align}
        Q_{\mathrm{g} ,\, j} & = \frac{\kappa_{j} \sigma_{\mathrm{D} ,\, \mathrm{g} ,\, j}}{\pi \mathrm{G} \, \Sigma_{\mathrm{g} ,\, j}} \label{equ:gaseous_toomre} \;,\\
        Q_{\mathrm{s} ,\, j} & = \frac{\kappa_{j} \sigma_{\mathrm{D} ,\, \mathrm{s} ,\, j}}{3.36 \mathrm{G} \, \Sigma_{\mathrm{s} ,\, j}} \label{equ:stellar_toomre} \;,
    \end{align}
\end{subequations}
where $Q > \num{1}$ for a stable disk.
Here we introduced for the gaseous and stellar disks, respectively, the surface densities $\Sigma_{\mathrm{g}} \,/\, \Sigma_{\mathrm{s}}$ and the velocity dispersions $\sigma_{\mathrm{D} ,\, \mathrm{g}} \,/\, \sigma_{\mathrm{D} ,\, \mathrm{s}}$.

We follow \citet[][but see also \citealp{vanderkruit2011a}]{bottema1993a} to calculate the velocity dispersion of the stars\footnote{Here we completely neglect the increase in velocity dispersion as a function of stellar age caused by interactions with giant molecular clouds or spiral waves, potentially resulting in a more stable disk.} as
\begin{equation}
    \sigma_{\mathrm{D} ,\, \mathrm{s} ,\, j} = \frac{v_{c ,\, j}}{2} \exp \bigg( -\frac{\langle w_{j} \rangle}{2w_{\mathrm{D}}} \bigg) \;,
    \label{equ:velocity_dispersion_stellar_disk}
\end{equation}
with circular velocity
\begin{equation}
    v_{c ,\, j} = \sqrt{ \langle w_{j} \rangle \frac{\partial \Phi_{\mathrm{tot}}}{\partial w} \bigg|_{\langle w_{j} \rangle } } \;.
    \label{equ:circular_velocity}
\end{equation}

We assume that the velocity dispersion of the cold gas equals the speed of sound of the interstellar medium, which correlates with the star formation surface density, i.e.\
\begin{equation}
    c_{\mathrm{s} ,\, \mathrm{cold} ,\, j} = \alpha_{\mathrm{cold}} + \beta_{\mathrm{cold}} \, \left( \frac{\Sigma_{\mathrm{SFR} ,\, j}}{\si{\Msun\per\kilo\pc\squared\per\yr}} \right)^{\gamma_{\mathrm{cold}}} \;,
    \label{equ:speed_of_sound_cold_gas}
\end{equation}
with free parameters $\alpha_{\mathrm{cold}}$, $\beta_{\mathrm{cold}}$, and $\gamma_{\mathrm{cold}}$.
This relationship shows significant scatter in observations across various redshifts \citep[see e.g.][and references therein]{lehnert2009a,genzel2011a,green2014a,krumholz2016a,zhou2017a,krumholz2018b,mai2024a}.
In our fiducial model we assume the parameter values $\alpha_{\mathrm{cold}} = 5 \, \textrm{km} \, \textrm{s}^{-1}$, $\beta_{\mathrm{cold}} = 20 \, \textrm{km} \, \textrm{s}^{-1}$, and $\gamma_{\mathrm{cold}} = 1 / 3$, i.e.\ a turbulence-dominated energy dissipation prescription for the cold gas \citep[][]{zhou2017a}.
We explore the impact of different values in \Cref{subsec:velocity_dispersion_of_the_cold_gas}.

It is well known that the gaseous and stellar disks interact dynamically \citep[e.g.][]{lin1966a,bertin1988a} and that, for example, the stability of a stellar disk may be impacted by even small amounts of gas.
To retain the same guidelines for the Toomre stability parameter in the prescription provided by \citet{kruijssen2012d} we follow the approach by \citet{romeo2011a} and compute an ``effective'' Toomre parameter as
\begin{equation}
    Q_{\mathrm{eff} ,\, j}^{-1} =
    \begin{dcases}
        \psi_{Q ,\, j} Q_{\mathrm{s} ,\, j}^{-1} + Q_{\mathrm{g} ,\, j}^{-1} & \mathrm{if} \; Q_{\mathrm{s} ,\, j} \geq Q_{\mathrm{g} ,\, j} \;,\\
        Q_{\mathrm{s} ,\, j}^{-1} + \psi_{Q ,\, j} Q_{\mathrm{g} ,\, j}^{-1} & \mathrm{otherwise} \;,
    \end{dcases}
    \label{equ:toomre}
\end{equation}
with the weighting factor
\begin{equation}
    \psi_{Q ,\, j} = 2\frac{\sigma_{\mathrm{D} ,\, \mathrm{s} ,\, j} \; \sigma_{\mathrm{D} ,\, \mathrm{g} ,\, j}}{\sigma_{\mathrm{D} ,\, \mathrm{s} ,\, j}^{2} + \sigma_{\mathrm{D} ,\, \mathrm{g} ,\, j}^{2}} \;.
    \label{equ:toomre_weight}
\end{equation}

\begin{figure*}
    \centering
    \includegraphics[width=\textwidth]{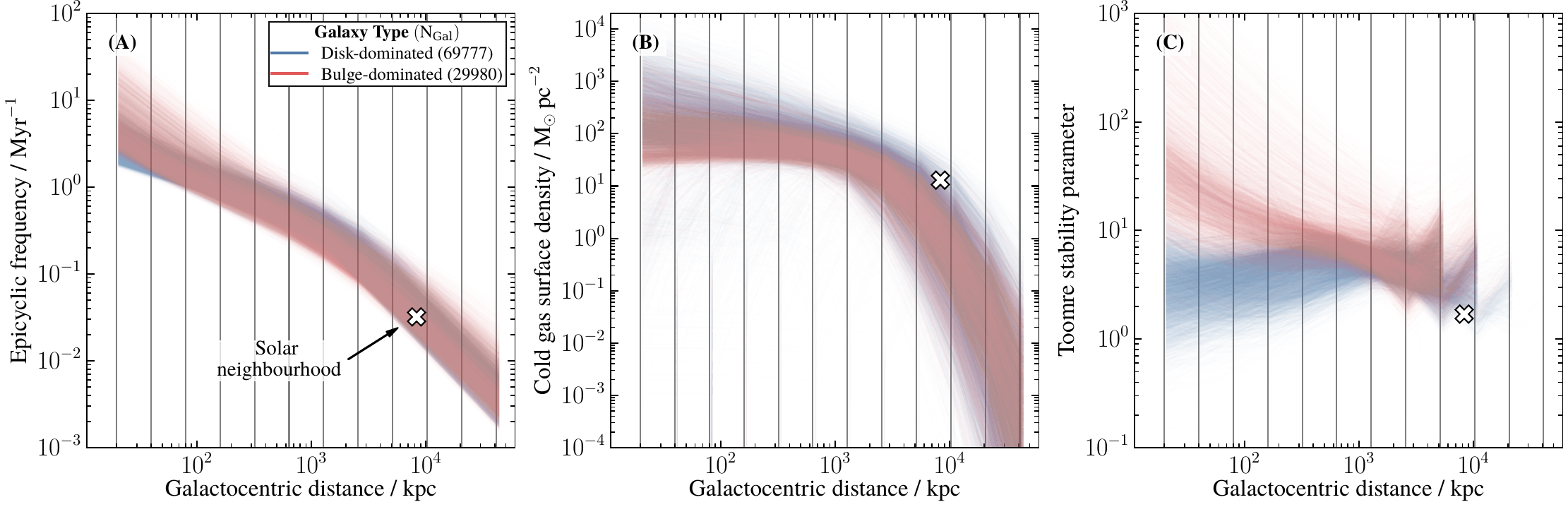}
    \caption{%
        Epicyclic frequency, cold gas surface mass density, and the Toomre stability parameter as a function of galactocentric distance for disk- (blue) and bulge-dominated (red) galaxies, defined as having a bulge-to-total stellar mass ratio of $B/T < 0.2$ and $B/T \geq 0.9$, respectively.
        We add for comparison the value of the solar neighbourhood: we calculate $\kappa_{\mathrm{D} ,\, \odot} \approx 0.046 \, \textrm{Myr}^{-1}$, as derived from the Oort constants $A = 15.6 \, \textrm{km} \, \textrm{s}^{-1} \, \textrm{kpc}^{-1}$ and $B = -15.8 \, \textrm{km} \, \textrm{s}^{-1} \, \textrm{kpc}^{-1}$ taken from \citet{guo2023a}; $\Sigma_{\mathrm{g} ,\, \odot} \approx 13 \, \textrm{M}_{\odot} \, \textrm{pc}^{-2}$ from \citet{flynn2006b}; and $Q_{\mathrm{eff} ,\, \odot} \approx 1.7$ \citep[][with $Q_{\mathrm{s} ,\, \odot} \approx 2.7$ and $Q_{\mathrm{g} ,\, \odot} \approx 1.5$]{binney2008a}, a typical value for disks \citep[e.g.][]{rafikov2001b,leroy2008a,feng2014a,westfall2014a}.
        Note that we do not calculate the Toomre stability parameter for annuli where it's surface density drops below $1 \, \textrm{M}_{\odot} \, \textrm{pc}^{-2}$ as we do not expect the formation of any star clusters at such low gaseous densities.
    }
    \label{fig:annuli_frequencies_surface_density_toomre}
\end{figure*}

We show in \Cref{fig:annuli_frequencies_surface_density_toomre} an overview of the cold gas surface mass density, the epicyclic frequency, and the Toomre stability parameter from running our model on tree-files of the Millennium simulation.
For comparison, we add the position of the solar neighbourhood.

The epicyclic frequency and cold gas surface density decrease at larger galactocentric distances.
The Toomre parameter shows a more complex behaviour, typically ranging between one and ten, except for the centre of elliptical galaxies and the largest distances where it increases to higher values.
The former effect is caused by the importance of the bulge component which is weaker in spirals.
At large radii the disks of low-mass galaxies are barely populated with gas and stars, which drive $Q_{\mathrm{eff}}$ to large values.
For the same reason the Toomre parameter drops for more massive galaxies, which is reflected by an increase in the disk's scale-length.

Overall we find good agreement between our simulated galaxies and the solar neighbourhood.

\subsubsection{Bound fraction of star formation}
\label{subsubsec:bound_fraction_of_star_formation}

The bound fraction of newly formed stars in star clusters is closely related to the cluster formation efficiency \citep[][]{bastian2008b,goddard2010a,adamo2011c,silva-villa2011a}, which takes into account its survival rate during the first few Myr.
Although important in many aspects, it is unclear how the bound fraction and cluster formation efficiency are related to the interstellar medium and star formation \citep[see][for a recent discussion on how feedback influences the cluster formation efficiency]{andersson2024a}.

As mentioned above, we follow the model outlined by \citet{kruijssen2012d} to estimate the bound fraction based on its epicyclic frequency, cold gas surface density, and Toomre stability parameter, all equated at the log-mean galactocentric distance.
We briefly highlight key aspects of the model and refer the interested reader to the original work for a more detailed description.

Note that we do not directly determine the bound fraction during the execution of the simulation.
Instead, to reduce the computational cost, we create lookup tables for a set of $\{\Sigma_{\mathrm{g}}, Q_{\mathrm{eff}}, \kappa\}$.
In total, we utilise $30$ lookup tables with varying values of $Q_{\mathrm{eff}}$ (between $0.5$ and $100$) and $500$ values of $\Sigma_{\mathrm{g}}$ and $\kappa$ each, resulting in $7.5 \times 10^{6}$ data points.
The simulation then determines the closest match in all three parameters and extracts the bound fraction from the table.

Following an extensive literature \citep[e.g.][]{padoan1997c,scalo1998a,ostriker2001a,kritsuk2007a,padoan2011a,kritsuk2017a,burkhart2018a} we assume that the density contrast of the interstellar medium follows a log-normal distribution of the form
\begin{equation}
    \mathrm{d}p_{j} = \frac{1}{\sqrt{2\pi} \varsigma_{\rho ,\, j}} \exp \bigg[ -\bigg( \frac{\ln \delta_{j} - \overline{\ln \delta_{j}}}{\sqrt{2} \varsigma_{\rho ,\, j}} \bigg)^{2} \bigg] \, \mathrm{d} (\ln \delta_{j}) \;,
    \label{equ:pdf_ism}
\end{equation}
with relative density $\delta_{j} = \rho / \rho_{\mathrm{ISM} ,\, j}$, $\overline{\ln \delta_{j}} = -0.5 \, \varsigma_{\rho ,\, j}^{2}$ \citep[e.g.][]{vazquez-semadeni1994a}, and standard deviation
\begin{equation}
    \varsigma_{\rho ,\, j} = \sqrt{\ln(1 + 3\gamma_{\rho}^{2}\mathcal{M}_{\mathrm{cold} ,\, j}^{2})} \;,
    \label{equ:ism_standard_deviation}
\end{equation}
where $\gamma_{\rho} \approx 0.5$ \citep{nordlund1999a,padoan2002a}.
The Mach number of the cold gas is related to the cold gas surface density, the epicyclic frequency, and the Toomre stability parameter via
\begin{equation}
    \mathcal{M}_{\mathrm{cold} ,\, j} = \sqrt{2} \phi_{\overline{\mathrm{P}} ,\, j}^{1/8} \frac{Q_{\mathrm{eff} ,\, j} \Sigma_{\mathrm{g} ,\, j}}{\kappa_{j}} \;,
    \label{equ:mach_number_cold_gas}
\end{equation}
where $\phi_{\overline{\mathrm{P}} ,\, j}$ is the ratio of the mean pressure of a gaseous cloud related to the pressure at its surface \citep{krumholz2005d} with typical values close to two in dense regions \citep[][]{heyer2004a,rosolowsky2005a,schuster2007a,colombo2014a}.
Notice that the parameter $\phi_{\overline{\mathrm{P}} ,\, j}$ is directly related to the fraction of cold gas contained in giant molecular clouds (GMCs) as $\phi_{\overline{\mathrm{P}} ,\, j} \approx 10 - 8 \times f_{\mathrm{GMC} ,\, j}$.
We assume that this fraction only depends on the cold gas surface density \citep[][]{krumholz2005d}, i.e.\
\begin{equation}
    f_{\mathrm{GMC} ,\, j} = \Big[ \num{1} + \num{250} \,/\, (\Sigma_{\mathrm{g} ,\, j} [\textrm{M}_{\odot} \, \textrm{pc}^{-2}])^{2} \Big]^{-1} \;.
    \label{equ:fraction_gmc}
\end{equation}

Next, we need to evaluate the minimum-value star formation efficiency.
If star formation occurs on the free-fall time scale, this efficiency can be expressed as a combination of the specific SFR, $\textrm{sSFR}_{\mathrm{ff}}$, and the ratio of the feedback time scale, $t_{\mathrm{fb}}$, to the free-fall time scale, $t_{\mathrm{ff}}$, i.e.\
\begin{equation}
    \epsilon_{\mathrm{ff} ,\, j} = \mathrm{sSFR}_{\mathrm{ff} ,\, j} \, t_{\mathrm{fb} ,\, j} \,/\, t_{\mathrm{ff} ,\, j} \;.
    \label{equ:sfe_ff}
\end{equation}
For a spherical symmetric and homogeneous mass distribution,
\begin{equation}
    t_{\mathrm{ff} ,\, j} = \sqrt{\frac{3\pi}{32 \mathrm{G} \, \rho_{\mathrm{ISM} ,\, j}}} \;,
    \label{equ:time-scale_ff}
\end{equation}
where we determine the density of the interstellar medium as
\begin{equation}
    \rho_{\mathrm{ISM} ,\, j} = \frac{\phi_{\mathrm{P}}}{2\pi \mathrm{G}} \bigg( \frac{\kappa_{j}}{Q_{\mathrm{eff} ,\, j}} \bigg)^{2} \;,
    \label{equ:ism_density}
\end{equation}
with $\phi_{\mathrm{P}} \approx \num{3}$ \citep[see Appendix~A of ][]{krumholz2005d}.

For the feedback time scale \citep{kruijssen2012d},
\begin{equation}
    t_{\mathrm{fb} ,\, j} = \frac{t_{\mathrm{SN}}}{2} \Bigg[ \num{1} + \sqrt{\num{1} + \frac{4\pi^{2} \mathrm{G}^{2} \; t_{\mathrm{ff} ,\, j}}{\phi_{\mathrm{fb}} \; \mathrm{sSFR}_{\mathrm{ff} ,\, j} \; t_{\mathrm{SN}}^{2}} \, \bigg( \frac{Q_{\mathrm{eff} ,\, j} \; \Sigma_{\mathrm{g} ,\, j}}{\kappa_{j}} \bigg)^{\num{2}}} \Bigg] \;,
    \label{equ:time-scale_fb}
\end{equation}
where $t_{\mathrm{SN}}$ is the time scale for the first supernovae, which we assume to be $3 \, \textrm{Myr}$, $\phi_{\mathrm{fb}} = 5.28 \times 10^{2} \, \textrm{pc}^{2} \textrm{Myr}^{-3}$ \citep{kruijssen2012d}, and
\begin{equation}
    \frac{\mathrm{sSFR}_{\mathrm{ff} ,\, j}}{0.13} = 1 + \erf \Bigg[ \frac{\varsigma_{\rho ,\, j}^{2} - \ln \big( 0.68 \, \alpha_{\mathrm{vir}}^{2} \, \mathcal{M}_{\mathrm{cold} ,\, j}^{4} \big)}{2^{3/2} \varsigma_{\rho ,\, j}} \Bigg] \;,
    \label{equ:ssfr_ff}
\end{equation}
with the virial parameter of GMCs $\alpha_{\mathrm{vir}}$ \citep[][]{larson1981a}.
It relates the mass, size, and velocity dispersion of a GMC and takes values between $10^{-1}$ and $10^{1}$ \citep[e.g.][]{myers1988b,bertoldi1992b,heyer2009a,dobbs2011b,hopkins2012c}.
We set $\alpha_{\mathrm{vir}} = 1.3$ as proposed by \citet{mckee2003a}.

In case star formation is less efficient than is assumed in \Cref{equ:sfe_ff} we take $\epsilon_{\mathrm{inc} ,\, j} = \epsilon_{\mathrm{ff} ,\, j} \times t_{\mathrm{ff} ,\, j} / 10 \, \textrm{Myr}$.
If star formation is more efficient we set an upper bound of $\epsilon_{\mathrm{inc} ,\, j} = \epsilon_{\mathrm{max}} = \num{0.5}$ \citep[][]{matzner2000a}.
The resulting effective star formation efficiency is the minimum of the above values,
\begin{equation}
    \epsilon_{j} = \min (\epsilon_{\mathrm{ff} ,\, j} \,,\, \epsilon_{\mathrm{inc} ,\, j} \,,\, \epsilon_{\mathrm{max}}) \,.
    \label{equ:minimum_efficiency}
\end{equation}
Finally, the bound fraction of star formation can be computed by the normalised integral of the probability density function of the contrast of the interstellar medium combined with the minimum star formation efficiency,
\begin{equation}
    f_{\mathrm{bound} ,\, j} = \epsilon_{\mathrm{max}}^{\num{-1}} \frac{\int_{-\infty}^{\infty} \mathrm{d}\delta_{j} \, \epsilon_{j}^{2}(\delta_{j}) \, \delta_{j} (\mathrm{d}p_{j} / \mathrm{d}\delta_{j})}{\int_{-\infty}^{\infty} \mathrm{d}\delta_{j} \, \epsilon_{j}(\delta_{j}) \, \delta_{j} (\mathrm{d}p_{j} / \mathrm{d}\delta_{j})} \;.
    \label{equ:bound_fraction}
\end{equation}

\begin{figure}
    \centering
    \includegraphics[width=\columnwidth]{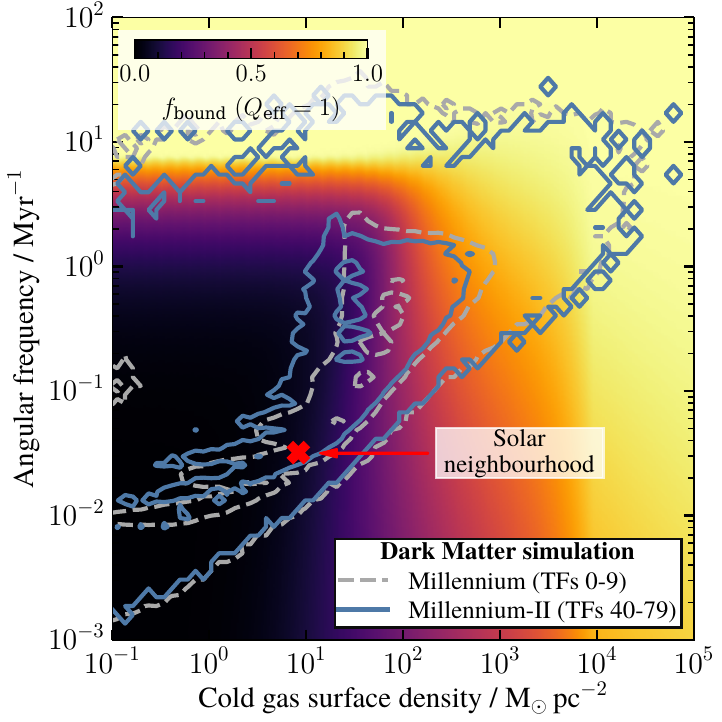}
    \caption{%
        Bound fraction of star formation, evaluated for $Q_{\mathrm{eff}} = 1.0$, as a function of angular frequency and cold gas surface density.
        Blue solid and gray dashed contours give the smoothed distribution (with standard deviation of $1 \, \textrm{dex}$) of all annuli of all galaxies running {\lgalaxies} tree-files 0-9 and 40-79 on the Millennium and Millennium-II simulations, respectively.
        The location of the solar neighbourhood (see \Cref{fig:annuli_frequencies_surface_density_toomre} for details) is marked with a cross.
    }
    \label{fig:cluster_bound_fraction}
\end{figure}

\Cref{fig:cluster_bound_fraction} shows the bound fraction for different environments with $Q_{\mathrm{eff}} = 0.5$ for the background.
The $z = 0$ distribution of annuli from running {\lgalaxies} on halos identified in the Millennium and Millennium-II simulations is added on top and reveals a large range bound star formation that depends on the location within the galaxy:
the innermost regions feature high surface densities and epicyclic frequencies (\textit{c.f.}~\Cref{fig:annuli_frequencies_surface_density_toomre}) and have high bound fractions approaching one.
In contrast, the outermost regions of galaxies have low surface densities and epicyclic frequencies, prohibiting the formation of bound structures.
As a consequence, this result already predicts that massive star clusters at distances a few times the disk's scale-length likely originate from accreted satellite galaxies, assuming that heating processes within their host galaxy are insignificant.

While not shown in \Cref{fig:cluster_bound_fraction}, we find a decrease in bound fraction for an increasing Toomre stability parameter value when keeping the epicyclic frequency and cold gas surface density constant.
This is related to the specific SFR, $\mathrm{sSFR}_{\mathrm{ff}}$, which decreases for an increasing Toomre parameter as per \Cref{equ:mach_number_cold_gas,equ:ssfr_ff}, and because the gas disk becomes more stable with an increase in $Q_{\mathrm{eff}}$.
As a result the mid-plane density of the interstellar medium decreases, which, in turn, decreases the star formation efficiency and, thus, the bound fraction.

\subsection{Initial star cluster properties}
\label{subsec:initial_star_cluster_properties}

Here we detail the initial properties of star clusters that form in galaxy disks.
Note that we deliberately exclude star clusters that form in the innermost ring with a radius of approximately $30 \, \textrm{pc}$.
These star clusters will either be disrupted quickly or merge to form a nuclear star cluster, which will be the subject of future work.

Another computational limit is the number of star clusters whose properties will be tracked over time.
In this work, we focus on the most massive star clusters that survive until $z = 0$ and decide to completely ignore star clusters below $10^{3} \ \textrm{M}_{\odot}$ (see below).
In the simulation we track the properties of the 1000 most massive star clusters in the disk and halo, respectively, resulting in up to 2000 objects per galaxy.
Although somewhat arbitrary, these numbers compare to the total number of star clusters of M{\,}31 \citep[potentially more than 1000; e.g.][]{barmby2001c,huxor2014a} and is much larger than the number of known globular clusters for the Milky Way \citep[about 200; see e.g.][and references therein]{minniti2017d,garro2024a}.

\subsubsection{Stellar mass}
\label{subsubsec:stellar_mass}

The mass of each star cluster is a random realisation of the underlying cluster initial mass function (CIMF), which we assume to follow a classical power-law distribution that is truncated at the upper mass end with an environmentally-dependent mass-scale, i.e.\
\begin{equation}
    \xi_{j}(m_{\mathrm{c}}) \propto m_{\mathrm{c}}^{\alpha_{\mathrm{CIMF}}} \, \exp ( -m_{\mathrm{c}} \,/\, m_{\mathrm{cl} ,\, \mathrm{max} ,\, j} ) \;.
    \label{equ:cimf_env}
\end{equation}
We assume here that $\alpha_{\mathrm{CIMF}} = -2$, motivated by observational studies of young star clusters in nearby galaxies \citep[e.g.][]{zhang1999a,bik2003a,hunter2003a,mccrady2007a,portegies-zwart2010b,emig2020b,levy2024a}.
The truncation mass-scale is a product of the star formation efficiency $\epsilon_{\mathrm{cloud}}$, the bound fraction of star formation $f_{\mathrm{bound}}$, the Toomre mass $m_{\mathrm{T}}$, and the fraction of molecular gas that is critical to undergo gravitational collapse $f_{\mathrm{coll}}$ \citep[][but see also \citealp{kruijssen2014c} and \citealp{reina-campos2022b}]{reina-campos2017a}, i.e.\
\begin{equation}
    m_{\mathrm{cl} ,\, \mathrm{max} ,\, j} = \epsilon_{\mathrm{cloud}} \, f_{\mathrm{bound} ,\, j} \, m_{\mathrm{T} ,\, j} \, f_{\mathrm{coll} ,\, j} \;,
    \label{equ:cimf_truncation}
\end{equation}
resulting in a typical value of the order of $10^{5} \, \textrm{M}_{\odot}$ at lower redshifts and up to $10^{9} \, \textrm{M}_{\odot}$ in extreme cases.
We assume
\begin{equation}
    m_{\mathrm{T} ,\, j} = 4\pi^{5} \mathrm{G}^{2} \, \Sigma_{\mathrm{g} ,\, j}^{3} \, Q_{\mathrm{eff} ,\, j}^{4} \,/\, \kappa_{j}^{4} \;,
    \label{equ:toomre_mass}
\end{equation}
and
\begin{equation}
    f_{\mathrm{coll} ,\, j} = \min ( 1 ,\, t_{\mathrm{fb} ,\, j} \,/\, t_{\mathrm{ff} ,\, \mathrm{2D} ,\, j} )^{4} \;,
    \label{equ:collision_fraction}
\end{equation}
for the Toomre mass and collapse fraction, respectively.
The two-dimensional free-fall time scale is $t_{\mathrm{ff} ,\, \mathrm{2D} ,\, j} = \sqrt{2\pi} / \kappa_{j}$.

In the above equations we assume a constant value of $\epsilon_{\mathrm{cloud}} = 0.1$ for star formation within a GMC, motivated by numerical results \citep[e.g.][]{oklopcic2017a,chevance2020a}.
Notice that this value is potentially smaller than the assumed star formation efficiency in the determination of the bound fraction in \Cref{equ:minimum_efficiency}.
A higher efficiency would result in a higher upper truncation mass-scale of the CIMF and could result in the formation of more massive star clusters.
However, as we show in \Cref{subsec:cluster_initial_mass_function} the main results of our work do not significantly change when assuming no upper truncation mass-scale but a pure power-law function instead.

For computational efficiency of the code, we consider only star clusters with initial masses in the range of $10^{4}$ to $10^{8} \, \textrm{M}_{\odot}$.
At the same time, we assume that the minimum star cluster mass that could, in principle, form is $10^{2} \, \textrm{M}_{\odot}$, meaning that we do not fully sample the CIMF.
Therefore, we only sample a total star cluster mass of $f_{\mathrm{sample}} \; f_{\mathrm{bound}} \; \mathrm{SFR} \; \delta t$ where
\begin{equation}
    f_{\mathrm{sample}} = \frac{\int_{10^{4} \, \textrm{M}_{\odot}}^{10^{8} \, \textrm{M}_{\odot}} m_{\mathrm{c}} \, \xi(m_{\mathrm{c}}) \, \mathrm{d}m_{\mathrm{c}}}{\int_{10^{2} \, \textrm{M}_{\odot}}^{10^{8} \, \textrm{M}_{\odot}} m_{\mathrm{c}} \, \xi(m_{\mathrm{c}}) \, \mathrm{d}m_{\mathrm{c}}} \;.
    \label{equ:mass_ratio}
\end{equation}
In the case of a power-law CIMF with index $-2$ this ratio is two-thirds, i.e.\ one-third of the total mass in star clusters is contained in objects below $10^{4} \, \textrm{M}_{\odot}$.
For the more complex case of \Cref{equ:cimf_env}, we pre-compute the integral numerically for $21$ equally-spaced values between $10^{3}$ and $10^{9} \, \textrm{M}_{\odot}$, resulting in, for example, $f_{\mathrm{sample}} \approx 2.3 \times 10^{-6}$ for $m_{\mathrm{cl} ,\, \mathrm{max}} = 10^{3} \, \textrm{M}_{\odot}$.
Afterwards, we randomly sample the CIMF with $m_{\mathrm{cl} ,\, \mathrm{max}}$ that agrees best with the computed value of \Cref{equ:cimf_truncation}.

Finally, to reduce computational cost we utilise lookup tables for initial star cluster masses.
For simplicity, we generated $21$ lookup tables for different values of $m_{\mathrm{cl} ,\, \mathrm{max}}$, equally spaced between $10^{3}$ and $10^{9} \, \textrm{M}_{\odot}$, with $10^{7}$ data points each.

\subsubsection{Initial galactocentric distances}
\label{subsubsec:initial_galactocentric_distances}

For each annulus we assume that the galactocentric distribution of initial values is uniform and independent of other star cluster parameters.

\subsubsection{Half-mass radius}
\label{subsubsec:cluster_size}

The physical processes that govern the distribution of the initial half-mass radius of star clusters is still unknown, and many theoretically motivated and observational-based prescriptions seem to fail to reproduce the distribution at $z = 0$ in nearby galaxies \citep[e.g.][]{reina-campos2023b}.
For that reason we adopt a simplified prescription by using a constant initial value of $r_{\mathrm{c} ,\, \mathrm{h}} = 1.0 \, \textrm{pc}$ for all clusters, independent of mass and redshift of formation.
We explore the impact of a different value and more complex prescriptions in \Cref{app:subsec:initial_half-mass_radius}.

\subsubsection{Tidal radius}
\label{subsubsec:tidal_radius}

The half-mass sizes of star clusters typically increase over time (\textit{c.f.}~\Cref{subsubsec:radial_expansion}) and are limited to the tidal radius where the gravitational acceleration of the cluster equals the tidal acceleration.
The tidal field is directly related to the local gravitational potential and the tidal radius is related to the first eigenvector of the diagonalised tidal tensor.
Assuming circular orbits and a mass concentration in the galaxy centre we apply the definition of \citet[][but see also \citealp{renaud2011a,renaud2018b}]{king1962a} and calculate
\begin{equation}
    r_{\mathrm{c} ,\, t} = \bigg[ \frac{\mathrm{G} \, m_{\mathrm{c}}}{(\Omega^{2} - \partial^{2} \Phi_{\mathrm{tot}} / \partial w^{2})_{w_{\mathrm{c}}}} \bigg]^{1/3} \;,
    \label{equ:cluster_radius_tidal}
\end{equation}
where $\Omega$ equals the angular frequency equated at the galactocentric distance of the star cluster.
More generally, for each ring $j$,
\begin{equation}
    \Omega_{j} = \sqrt{ \frac{1}{\langle w_{j} \rangle} \frac{\partial \Phi_{\mathrm{tot}}}{\partial w} \bigg|_{\langle w_{j} \rangle} } \;.
    \label{equ:angular_frequency}
\end{equation}

Note that this prescription ignores the tidal effect of nearby baryonic over-densities, such as star forming regions or clouds, which may be dominant over the global galactic field.
Other works that use hydrodynamical approaches \citep[such as][]{reina-campos2022b} determine a local tidal tensor from neighbouring cells, which is not possible in our model.
We discuss this issue in \Cref{subsec:caveats}.

\subsubsection{Metallicity}
\label{subsubsec:metallicity}

For the metallicity (and individual elemental abundances of H, He, C, N, O, Ne, Mg, Si, S, Ca, Fe) of a star cluster we assume that it equals the metallicity of the cold gas in the ring it forms in, $Z_{\mathrm{c}} = Z_{\mathrm{g}}$, and that this value remains constant over the star cluster's lifetime.
The first assumption neglects any azimuthal variations of metallicity profiles which are known to exist in some galaxies due to asymmetric structures such as bars or spiral patters in the Milky Way \citep[e.g.][]{poggio2022a,spina2022a,filion2023a,hawkins2023b,hackshaw2024a} and other spiral galaxies \citep[e.g.][]{sanchez-menguiano2016a,ho2017a,sanchez-menguiano2018a,ho2019a,hwang2019a,metha2021a,metha2024a} but see \citet[][]{kreckel2019a} for counterexamples.
Nevertheless, despite lacking such an implementation we argue in \Cref{subsec:metallicity} that the metallicity distributions of our cluster populations are reasonable with respect to observational constraints.

\subsection{Star cluster evolution}
\label{subsec:star_cluster_evolution}

We detail in the next subsections the main processes that affect the evolution of star clusters: mass loss rates, expansion of the half-mass radius, and re-location.

\subsubsection{Mass-loss}
\label{subsubsec:mass-loss}

We consider three mechanisms for cluster mass-loss: stellar evolution, tidal stripping due to an expanding cluster, and tidal shocks from interactions with GMCs, i.e.\
\begin{equation}
    \frac{\mathrm{d}m_{\mathrm{c}}}{\mathrm{d}t} = \frac{\mathrm{d}m_{\mathrm{c}}}{\mathrm{d}t} \bigg|_{\mathrm{ev}} + \frac{\mathrm{d}m_{\mathrm{c}}}{\mathrm{d}t} \bigg|_{\mathrm{rlx}} + \frac{\mathrm{d}m_{\mathrm{c}}}{\mathrm{d}t} \bigg|_{\mathrm{sh}} \;.
    \label{equ:total_mass_loss}
\end{equation}

\paragraph{Stellar evolution}

To take into account cluster mass-loss from stellar evolution we assume that cluster members represent a random realisation of a single stellar population assuming a Chabrier IMF \citep{chabrier2003a}, resulting in varying expected lifetimes.
Then, for an individual $10^{5} \, \textrm{M}_{\odot}$ star cluster we utilise the  ``Stochastically Lighting Up Galaxies'' library \citep{da-silva2012a,da-silva2014c,krumholz2015b} with non-rotating ``Geneva'' 2013 stellar tracks \citep{schaller1992a,meynet1994a,ekstroem2012a,georgy2013b} and fit a linear relationship to the retained mass as a function of time.
The resulting mass-loss rate at time $t$ in [yr] is
\begin{equation}
    \frac{\mathrm{d}m_{\mathrm{c}}}{\mathrm{d}t} \bigg|_{\mathrm{ev}} = -\frac{0.13}{\ln(10)} \frac{m_{\mathrm{c}}^{\mathrm{init}}}{t},\, \; \mathrm{for} \; t \geq t_{\mathrm{SN}} \;.
    \label{equ:mass-loss_stellar_evolution}
\end{equation}
We find that this relation holds, to first order, irrespective of cluster metallicity.
Note that the above determination does not take into account other mass-loss channels that we introduce below, which is why we update the ``initial'' star cluster mass, denoted here with $m_{\mathrm{c}}^{\mathrm{init}}$, at each time-step.

\paragraph{Relaxation}

Multi-body encounters between stars within a cluster result in energy transfer between the individual bodies and can cause stars to either orbit the star cluster's centre at larger radii or leave the cluster completely in case its velocity exceeds the escape velocity.
For bound stars, if the new orbit crosses the tidal radius, the star can be stripped from the cluster, resulting in an effective mass loss.

We consider this effect by following an extensive literature \citep[e.g.][]{spitzer1940a,henon1961a,spitzer1987a,lamers2005a} and set
\begin{equation}
    \frac{\mathrm{d} m_{\mathrm{c}}}{\mathrm{d} t} \bigg|_{\mathrm{rlx}} = -\xi_{\mathrm{rlx}} \frac{m_{\mathrm{c}}}{\tau_{\mathrm{rlx}}} \;,
    \label{equ:mass-loss_relaxation}
\end{equation}
with a relaxation time scale
\begin{equation}
    \tau_{\mathrm{rlx}} = 0.138 \frac{\sqrt{N}}{\ln (\gamma_{\mathrm{rlx}} N)} \sqrt{\frac{r_{\mathrm{c} ,\, h}^{3}}{\mathrm{G} \, \langle m_{\star}\rangle}} \;.
    \label{equ:time_relaxation}
\end{equation}
Here $\langle m_{\star} \rangle$ is the average stellar mass of the star cluster, $N = m_{\mathrm{c}} \,/\, \langle m_{\star} \rangle$, and $0.07 \lesssim \gamma_{\mathrm{rlx}} \lesssim 0.14$ \citep{giersz1994a}.
We assume a \citet{chabrier2003a} initial stellar mass function, such that $\langle m_{\star} \rangle = 0.42 \, \textrm{M}_{\odot}$ and $\gamma_{\mathrm{rlx}} = 0.11$.
Finally, we choose $\xi_{\mathrm{rlx}} = 0.08$ as suggested in the literature \citep[e.g.][]{henon1961a,gieles2011b,gieles2016a} for equal-mass cluster members, avoiding a proper treatment of the star cluster's direct tidal environment \citep{alexander2012b}.

\paragraph{Tidal shocks}

When a star cluster is located within the thin disk, i.e.\ has not been accreted during a galaxy merger event, it frequently interacts with GMCs if the fraction of cold gas bound within clouds is high.
Depending on the impact parameter between the interaction the GMC can inject a significant amount of energy into the star cluster resulting in an increase in velocity dispersion and causing a fraction of the stars to escape the cluster as their velocity exceeds the cluster's escape velocity.

To model this effect we approximate a cluster's internal energy by assuming that it follows a Plummer-like density profile \citep{plummer1911a}.
Following the equations outlined in \citet{kruijssen2012d} we set
\begin{equation}
    \frac{\mathrm{d} m_{\mathrm{c}}}{\mathrm{d} t} \bigg|_{\mathrm{sh}} = -\frac{m_{\mathrm{c}}}{\tau_{\mathrm{sh}}} \;,
    \label{equ:mass-loss_shocks}
\end{equation}
with
\begin{equation}
    \tau_{\mathrm{sh}} = \frac{3 \sqrt{\pi}}{8 \sqrt{2^{2/3} - 1}}\frac{\mathrm{G}}{g \phi_{\mathrm{sh}} \phi_{\mathrm{ad}} \phi_{\mathrm{P}} f_{\mathrm{sh}} f_{\mathrm{GMC}}} \, \Bigg[ \frac{Q_{\mathrm{eff} ,\, j}}{\kappa_{j}} \Bigg]_{w_{\mathrm{c}}}^{3} \frac{m_{\mathrm{c}}}{r_{\mathrm{c} ,\, h}^{3}} \;,
    \label{equ:time_shocks}
\end{equation}
with $g = 1.5$, $\phi_{\mathrm{sh}} = 2.8$, $f_{\mathrm{sh}} = 3$ and $\phi_{\mathrm{ad}} = \exp (-0.062)$.
We use \Cref{equ:fraction_gmc} to equate $f_{\mathrm{GMC}}$ for the annuli corresponding to the star cluster's current galactocentric distance.

Note that only star clusters in galaxy disks are affected by tidal shocks due to encounters with GMCs, i.e.\ we set $\mathrm{d} m_{\mathrm{c}} / \mathrm{d} t |_{\mathrm{sh}} = 0$ for star clusters in a galaxy's halo.
Our implementation also does not couple the time scale of tidal shocks to the strength of the local tidal field tensor \citep[see e.g.][for details]{alexander2014a,reina-campos2022b} because we do not attempt to model GMCs.
However, because {\lgalaxies} already separates the atomic from the molecular gas in each annulus \citep[see][but also \citealp{yates2024a}]{henriques2020a,yates2021a} future efforts may implement clouds and improve on the current prescription.

\subsubsection{Radial expansion}
\label{subsubsec:radial_expansion}

Star clusters expand adiabatically due to contributions from mass-loss from two-body interactions and tidal shocks \citep[e.g.][]{reina-campos2022b}.
This results in a radial expansion of the form
\begin{equation}
    \frac{\mathrm{d} r_{\mathrm{c} ,\, h}}{\mathrm{d} t} = \Bigg[ \big( 2 - f_{\mathrm{sh}}^{-1} \big) \frac{\mathrm{d} m_{\mathrm{c}}}{\mathrm{d} t} \bigg|_{\mathrm{sh}} + \big( 2 - \zeta \xi_{\mathrm{rlx}}^{-1} \big) \frac{\mathrm{d} m_{\mathrm{c}}}{\mathrm{d} t} \bigg|_{\mathrm{rlx}} \Bigg] \frac{r_{\mathrm{c} ,\, h}}{m_{\mathrm{c}}} \;,
    \label{equ:evolution_half-mass_radius}
\end{equation}
Following \citet{gieles2016a} we assume that $f_{\mathrm{sh}} = 3$ and following \citet{gieles2011b}
\begin{equation}
    \zeta \xi_{\mathrm{rlx}}^{-1} = \frac{\num{5}}{\num{3}} \Bigg[ \bigg( \frac{r_{\mathrm{c} ,\, h}}{r_{\mathrm{c} ,\, j}} \bigg)_{1} \frac{r_{\mathrm{c} ,\, j}}{r_{\mathrm{c} ,\, h}} \Bigg]^{3/2} \approx 0.092 \, \bigg( \frac{r_{\mathrm{c} ,\, t}}{r_{\mathrm{c} ,\, h}} \bigg)^{3/2} \;,
    \label{equ:evolution_half-mass_radius_zeta_xi}
\end{equation}
where we assumed in the approximation that the Jacobi radius equals the tidal radius and that $\big( r_{\mathrm{c} ,\, h} / r_{\mathrm{c} ,\, j} \big)_{1} \approx 0.145$ \citep{henon1961a}.
The particular choice of $f_{\mathrm{sh}}$ results in a contraction of a star cluster from encounters with GMCs as the mass-loss from shocks is negative.
A star cluster can only expand if the contribution of two-body relaxation dominates, which strictly requires $\zeta \xi_{\mathrm{rlx}}^{-1} > 2$ or $r_{\mathrm{h}} \lesssim 0.128 \, r_{\mathrm{t}}$.
This condition is met in low-density environments, such as a galaxy's halo, and we expect star clusters to contract in the inner parts of a galaxy's disk.

\subsubsection{Galactocentric migration}
\label{subsubsec:galactocentric_migration}

Without any external perturbations star clusters in a galaxy's disk and halo migrate towards the inner regions through dynamical friction caused by interactions with constituents of three components: (1) field stars belonging to the stellar halo and bulge, (2) the gaseous halo, and (3) the dark matter halo.
For the stellar and dark matter components, we follow the well-known \citet{chandrasekhar1943c} prescription and assume an isotropic velocity distribution function of the components as well as circular orbits of the star clusters.
As a consequence, a star cluster experiences a radial acceleration of
\begin{equation}
    \frac{\mathrm{d} v_{\mathrm{c}}}{\mathrm{d} t} = -4\pi \mathrm{G}^{2} \frac{m_{\mathrm{c}} \rho_{\mathrm{f}}}{v_{\mathrm{c}}^{2}} \ln \Lambda \bigg[ \erf (X) - \frac{2X}{\sqrt{\pi}} \exp \big( -X^{2} \big) \bigg] \;,
    \label{equ:dynamical_friction_1}
\end{equation}
where $X = v_{\mathrm{c}} \,/\, (2 \sigma_{\mathrm{f}})$ and $\rho_{\mathrm{f}}$ is the density of dark matter and stars in the halo evaluated at the position of the star cluster.
The circular velocity of each object is computed via \Cref{equ:circular_velocity}.
For the Coulomb logarithm, we follow \citet{binney2008a} and assume
\begin{equation}
    \ln \Lambda = \ln \bigg[ \frac{w_{\mathrm{c}}}{\max (r_{\mathrm{c} ,\, h} \,, \mathrm{G} \, m_{\mathrm{c}} \,/\, v_{\mathrm{typ}}^{2})} \bigg] \;,
    \label{equ:coulomb_logarithm}
\end{equation}
where we replaced the maximum impact parameter with the galactocentric distance of the star clusters and assume for the typical velocity parameters of the dark matter halo, $v_{\mathrm{typ}} = \sqrt{\mathrm{G} \, M_{\mathrm{vir}} / R_{\mathrm{vir}}}$.
For the velocity dispersion of halo stars we assume that their orbits are dominated by the underlying dark matter profile.
Based on the Jeans equation for an isotropic system \citet{zentner2003a} give the following approximation for the velocity dispersion,
\begin{equation}
    \sigma_{\mathrm{H} ,\, \mathrm{s}} = \frac{1.4393 \, x_{\mathrm{c}}^{0.354}}{1 + 1.1756 \, x_{\mathrm{c}}^{0.725}} \, v_{\mathrm{max}} \;,
    \label{equ:velocity_dispersion_halo_stars}
\end{equation}
with maximum circular velocity within the dark matter halo $v_{\mathrm{max}}$ that we introduced in \Cref{equ:dynamical_time} and $x_{\mathrm{c}} = w_{\mathrm{c}} / R_{\mathrm{vir}}$.

The radial acceleration caused by the hot gaseous component differs slightly from the one introduced in \Cref{equ:dynamical_friction_1}.
Following \citet{ostriker1999a} and \citet{escala2004a}, we take
\begin{equation}
    \frac{\mathrm{d} v_{\mathrm{c}}}{\mathrm{d} t} = -\frac{4\pi \mathrm{G}^{2} \, m_{\mathrm{c}} \rho_{\mathrm{H}}}{v_{\mathrm{c}}^{2}} g(\mathcal{M}_{\mathrm{hot}}) \;,
    \label{equ:dynamical_friction_2}
\end{equation}
where $\rho_{\mathrm{H}}$ is the density of the gaseous halo evaluated at the star clusters position and
\begin{equation}
    g(\mathcal{M}_{\mathrm{hot}}) = \frac{1}{2} \times
    \begin{cases}
        \ln \Big( \frac{1 + \mathcal{M}}{1 - \mathcal{M}} \Big) - 2 \mathcal{M} & \mathrm{if} \; \mathcal{M} < 1.0 \;, \\
        \ln (1 - \mathcal{M}^{-2}) + 2 \ln \Lambda & \mathrm{if} \; \mathcal{M} > 1.0 \;.
    \end{cases}
\end{equation}
The parameter $\ln \Lambda$ is the Coulomb-logarithm first introduced in \Cref{equ:coulomb_logarithm}.
We calculate the Mach number of a cluster within the hot gas halo as $\mathcal{M}_{\mathrm{hot}} = v_{\mathrm{c}} \,/\, c_{\mathrm{s} ,\, \mathrm{hot}}$ where \citep{tanaka2009a,choksi2017a}
\begin{equation}
    \frac{c_{\mathrm{s} ,\, \mathrm{hot}}}{[\si{\kilo\metre\per\second}]} = 1.8 \sqrt{1+z} \, \bigg( \frac{M_{\mathrm{vir}}}{10^{7} \, \textrm{M}_{\odot}} \bigg)^{1/3} \bigg( \frac{\Omega_{\mathrm{M ,\, 0}} \, h^{2}}{0.14} \bigg)^{1/6} \;.
    \label{equ:velocity_dispersion_hot_gas}
\end{equation}

\subsubsection{Re-distribution during galaxy mergers}
\label{subsubsec:re-distribution_during_galaxy_mergers}

Explanations for the observed presence of star clusters in galaxy halos generally favour (a) interactions with massive perturbers (such as other star clusters or GMCs) that heat the object from its birth-place in a disk, (b) star cluster formation in galaxy outskirts during tidal interactions, or (c) a direct re-distribution during galaxy mergers.
The lower tidal field and absence of tidal shocks may be an important ingredient in cluster survival as indicated by \citet{baumgardt2018b} who find that the Milky Way's halo hosts all Galactic star clusters older than approximately five Gyr.
Therefore, re-positioning of star clusters is an essential physical ingredient in simulating star cluster properties.

We consider two different scenarios based on the ratio of the baryonic masses of the two galaxies that participate in a merger.
Here we follow the prescription by {\lgalaxies} and assume that a major merger occurs if the mass ratio exceeds $q \geq 0.1$.
In this case, we assume that the disks of both galaxies are destroyed and that all star clusters from both galaxies are contained within the halo of the successor galaxy.
For minor galaxy mergers ($q < 0.1$) we assume that the star cluster population of the more massive galaxy remains unaffected and that all accreted star clusters migrate into the halo of their new host.

Generally speaking, the resulting baryonic distribution after galaxy mergers is sensitive to the initial conditions such as the galaxy mass ratio and their respective positions, orientations, and velocity vectors and magnitudes to each other.
We therefore make simplifying assumption for the new galactocentric distances of accreted star clusters.

In {\lgalaxies} a galaxy starts to disrupt once the baryonic mass components equal or exceed the remaining dark matter mass.
Once this condition is met, the model computes the following dynamical friction time-scale \citep[][]{binney2008a},
\begin{equation}
    \tau_{\mathrm{fric}} = \alpha_{\mathrm{fric}} \frac{V_{200} \, d_{\mathrm{sat}}^{2}}{\mathrm{G} M_{\mathrm{sat}} \, \ln \Lambda_{\mathrm{fric}}} \;,
    \label{equ:time-scale_dynamical_friction_galaxy}
\end{equation}
where $V_{200}$ is the velocity at distance $R_{200}$ from the centre of the central halo, $M_{\mathrm{sat}}$ the total (baryonic and dark matter) mass of the to-be accreted satellite, and $d_{\mathrm{sat}}$ the distance from the satellite galaxy to the central's centre.
The parameter $\alpha_{\mathrm{fric}} = \num{2.4}$ ensures that the bright-end of the $z=0$ galaxy luminosity function is recovered \citep[][]{de-lucia2007b}.
Finally, $\ln \Lambda_{\mathrm{fric}}$ is the Coloumb logarithm, set to $\ln \Lambda_{\mathrm{fric}} = \ln (1 + M_{200} / M_{\mathrm{sat}})$, where $M_{200}$ is the enclosed mass within $R_{200}$ of the central galaxy.

The model assumes that the satellite follows a circular trajectory, such that its distance from the centre of the central halo is reduced by a factor $\Delta t / \tau_{\mathrm{fric}}$ where $\Delta t$ is the time since $\tau_{\mathrm{fric}}$ was first computed, incremented according to the time-step of the simulation.
Once $\Delta t \geq \tau_{\mathrm{fric}}$ the satellite is assumed to be fully accreted.

In our approach we make use of these two time-scales ($\tau_{\mathrm{fric}} = 0$ and $\tau_{\mathrm{fric}} = \Delta t$) by tracking the distance of the satellite between both times.\footnote{Note that the final distance, once $\tau_{\mathrm{fric}} = \Delta t$ typically is greater than zero (but often smaller than \SI{1}{\kilo\pc}) due to the invalidity of the assumption of circular motion of the satellite around the central halo.}
At the same time, once the code computes $\tau_{\mathrm{fric}}$, we determine the radius of the galaxy from the condition that the baryonic mass of the satellite galaxy matches the dark matter mass, under the assumption that the density profile of the dark matter within this radius remains unchanged from the stripping of the outer parts.
We find this radius ($w_{\mathrm{max}}$) by solving numerically the following equation,
\begin{equation}
    \frac{M_{\mathrm{Bary}}}{M_{\mathrm{DM} ,\, \mathrm{0}}} = \frac{\ln(1 + c_{\mathrm{vir}} \, w_{\mathrm{max}} / R_{\mathrm{vir}}) - c_{\mathrm{vir}} / (R_{\mathrm{vir}} / w_{\mathrm{max}} + c_{\mathrm{vir}})}{\ln(1 + c_{\mathrm{vir}}) - c_{\mathrm{vir}} / (1 + c_{\mathrm{vir}})} \;,
    \label{equ:calculation_wmax}
\end{equation}
where $M_{\mathrm{Bary}}$ is the sum of the gaseous and stellar disks, the bulge, and central black hole, and $M_{\mathrm{DM} ,\, 0}$ the dark matter mass of the satellite prior to in-fall.

For the second distance we assume that the galaxy is completely disrupted once $\tau_{\mathrm{fric}} = \Delta t$ where the central parts of the satellite get stripped, i.e.\ $w_{\mathrm{min}} = 0$.
We now relate the internal distances $w_{\mathrm{min}}$ and $w_{\mathrm{max}}$ to the distances of the satellite during disruption $d_{\mathrm{min}}$ and $d_{\mathrm{max}}$ and the position of a star cluster in the satellite at position $w_{\mathrm{c}}^{\mathrm{old}}$ to compute the new galactocentric distance $w_{\mathrm{c}}^{\mathrm{new}}$ within the central halo via
\begin{equation}
    w_{\mathrm{c}}^{\mathrm{new}} = \bigg( \frac{w_{\mathrm{c}}^{\mathrm{old}}}{w_{\mathrm{max}} - w_{\mathrm{min}}} \bigg)^{\alpha_{w}} (d_{\mathrm{max}} - d_{\mathrm{min}}) + d_{\mathrm{min}} \;,
    \label{equ:galactocentric_distance_mergers}
\end{equation}
where we assume in our fiducial model that $\alpha_{w} = 4$.
Thus, this prescription assumes that the most distant but still bound star clusters relocate to a distance $d_{\mathrm{max}}$ and the most central ones to $d_{\mathrm{min}}$.
Star clusters with $w_{\mathrm{c}}^{\mathrm{old}} > w_{\mathrm{max}}$ are stripped from the satellite halo before the dynamical friction time-scale is computed in the simulation.
Due to the large distances involved (typically larger than \SI{1}{\mega\pc}) we assume that these star clusters become unbound from the central galaxy.
We do not track their subsequent evolution.

Note our assumption that all star clusters survive the tidal shock experienced during galaxy mergers, i.e.\ resulting in a ``survival fraction'' of unity \citep[see][for a different approach]{kruijssen2012b,de-lucia2024a}.
Furthermore, we do not add another mass-loss term for this scenario.

\section{Results}
\label{sec:results}

In this work we focus on basic properties of the star cluster populations at $z = 0$.
In the following sections, we refer to a star cluster as young (old) if its age is $\tau_{\mathrm{c}} < 0.3 \, \textrm{Gyr}$ ($\tau_{\mathrm{c}} \geq 6 \, \textrm{Gyr}$), unless specified otherwise.
These age cuts are used throughout the analysis and while they are somewhat arbitrary, we chose them to be able to better compare to both observational results and predictions by other simulations.

Disk-dominated (``spiral'') galaxies are assumed to be the ones with a bulge-to-total stellar mass ratio of $B/T < 0.2$ whereas bulge-dominated (``elliptical'') systems have $B/T \geq 0.7$.\footnote{{\lgalaxies} does not model any spiral-wave patterns in galaxies, which is why we prefer the term ``disk-dominated'' over ``spiral''.}

We focus our analysis on the output of running the model on Millennium \citep[][]{springel2005c} tree-files $0$-$9$ (out of $512$ total) that contain $118{\,}558$ galaxies at $z = 0$ and provide us with a representative sub-sample.
The model performs similarly when running on Millennium-II \citep[][]{boylan-kolchin2009a} tree-files $40$-$79$ (a representative sub-sample; out of $512$), going down to lower galaxy stellar masses, and we will detail a brief comparison in each subsection in case of differences.

\subsection{$M_{V}$-SFR relationship}
\label{subsec:mv-sfr_relationship}

A first test for our model is to reproduce the empirical relationship between the absolute $V$-band magnitude of the brightest young star cluster (local quantity within a galaxy) versus the host galaxy's SFR (global quantity) within nearby disk-dominated galaxies \citep[e.g.][]{larsen2002e,bastian2008b,larsen2010a}.
Since this relation is not used as input for the simulation it serves as both a check of the models capabilities and a test to explore secondary correlations with other third quantities.

For the simulated data, as we store star cluster masses, we need to convert to the Johnson-Cousins $V$-band.
We perform the conversion by using the Python version \citep{johnson2023a} of the ``Flexible Stellar Population Synthesis'' code \citep[][]{conroy2009c,conroy2010a,conroy2010d,conroy2010g} that uses as input the metallicity and age of a stellar population and yields an absolute magnitude in selected filterbands.
Furthermore, for the computation, we assume that all star clusters are composed of a single stellar population that follows a \citet{chabrier2003a} stellar initial mass function.
To compare to observations we compile the data from \citet{johnson2000b,larsen2002e,rafelski2005a,bastian2008b,annibali2009a,goddard2010a,adamo2011c,annibali2011b,pasquali2011a,silva-villa2011a,cook2012a,ryon2014a,whitmore2014a,adamo2015b,lim2015a,cook2023a} for a local sample of galaxies.
Additionally, we use data from \citet{maschmann2024b,thilker2025a}, representing the results for PHANGS galaxies \citep[see][for details]{lee2022a,lee2023a}.
For that data set specifically, we use stellar mass estimates from \citet{emsellem2022a} and SFR estimates from \citet{sun2022a}.
Otherwise, galaxy stellar masses are taken from the $50 \, \textrm{Mpc}$ catalogue provided by \citet{ohlson2023a}.

\begin{figure*}
    \centering
    \includegraphics[width=\textwidth]{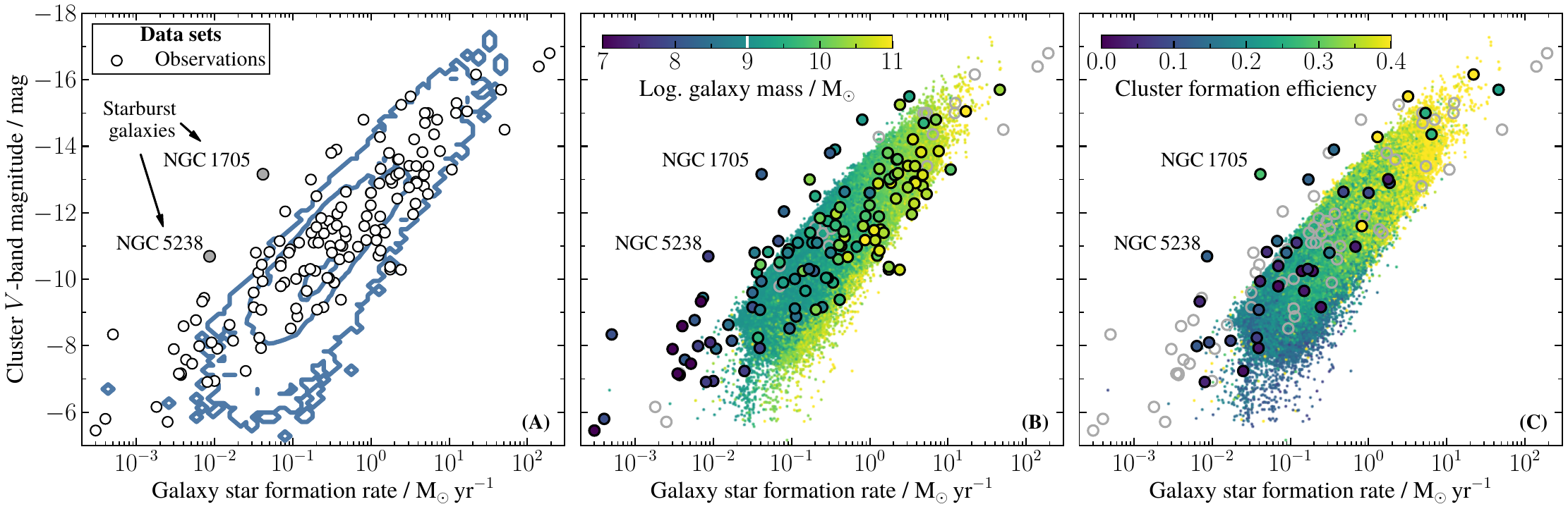}
    \caption{%
        Absolute $V$-band magnitude of the youngest and most massive star cluster versus the galaxy-averaged star formation rate.
        The galaxy sample is limited to disk-dominated galaxies that have a bulge-to-total stellar mass ratio of $B/T < 0.2$.
        We compare our results to various observations of nearby disk-dominated galaxies (see main text for details).
        For both the simulated data and the observations, we set an age cut of $\tau_{\mathrm{c}} \leq 0.3 \, \textrm{Gyr}$ on the star clusters.
        \textit{Panel A:} Full observational and simulated data samples.
        For the simulated data, we show the 1-, 2-, and 3-$\sigma$ intervals using blue solid lines.
        \textit{Panel B:} Same as in the first panel but colour-coding all data point by the host galaxy's stellar mass.
        If no stellar mass estimate is available for observational data points, we show them with gray symbols.
        \textit{Panel C:} Same as the central panel but colour-coding the data points by the cluster formation efficiency, which is a combination of the bound fraction of star formation and the ``cruel cradle effect'' \citep{kruijssen2012b,kruijssen2012d} that takes the interaction of a proto-star cluster with its natal environment and nearby giant molecular clouds into account.
        Note that the two outliers, NGC{\,}1705 and NGC{\,}5238, are starburst galaxies and that their massive star clusters were previously classified as nuclear star clusters \citep{pechetti2020a,hoyer2021a}.
        Nuclear star clusters often exhibit complex formation histories \citep[e.g.][]{spengler2017a,kacharov2018a,fahrion2021a} and cannot easily be compared to our simulated star clusters.
    }
    \label{fig:cluster_mv_sfr}
\end{figure*}

We show the resulting parameter space for young massive star clusters in disk-dominated galaxies evaluated at $z=0$ in \Cref{fig:cluster_mv_sfr}.
Our results show an increasing star cluster mass with increasing SFR, in qualitative agreement with the observations, although with a slightly steeper slope.
To quantify the level of (dis-)agreement we perform a linear fit to the data sample and obtain uncertainties through $10{\,}000$ Monte Carlo iterations.
The resulting slope values are $\alpha_{\mathrm{sim}} = -2.466^{+0.006}_{-0.006}$ for the simulated data ($N_{\mathrm{gal}} = 66292$) and $\alpha_{\mathrm{obs}} = -2.02^{+0.09}_{-0.09}$ for the observations ($N_{\mathrm{gal}} = 103$).
Despite small differences in the slope, most of our star clusters are located in galaxies with SFRs of the order of $10^{-1} \, \textrm{M}_{\odot} \, \textrm{yr}^{-1}$ and have absolute $V$-band magnitudes of $M_{V} \approx -11$, which is in excellent agreement with observations.

When running the model on Millennium-II data we find a better agreement for the lowest-mass (but still the most massive and young) star clusters towards lower galaxy SFRs, which is because of the higher mass resolution offered by the simulation.
We observe that the slope values of the relationship changes: applying a galaxy mass cut of $10^{9.5} \, \textrm{M}_{\odot}$ results in a slope of $\alpha_{\mathrm{sim} ,\, \mathrm{mrii}} = -2.02^{+0.03}_{-0.03}$.
This value matches the results for the observational data set.

We identify a secondary dependence on galaxy stellar mass in panel (B) where the slope of the relationship is steeper for massive galaxies:
when constraining the galaxy sample to stellar masses above $10^{10} \, \textrm{M}_{\odot}$ the slope steepens to $\alpha_{\mathrm{sim}} \approx -3.16$ ($N_{\mathrm{gal}} = 14091$).\footnote{Note that the secondary dependence on galaxy mass is also present in the {\emosaics} model \citep[Figure~10 in ][]{pfeffer2019b}, however, they only include $153$ galaxies and have an additional $y$-axis offset of $0.5$ to $1 \, \textrm{dex}$.}
This trend is mainly related to the upper truncation mass-scale for the CIMF, which decreases due to a decreasing Toomre mass and bound fraction.
Galaxies with stellar masses below $10^{9.5} \, \textrm{M}_{\odot}$ have comparable $Q_{\mathrm{D} ,\, \mathrm{g}}$ values to more massive galaxies but their $Q_{\mathrm{D} ,\, \mathrm{s}}$ values are larger because of their disk mass is dominated by gas and not stars and $Q_{\mathrm{D} ,\, i} \propto \Sigma_{\mathrm{D} ,\, i}^{-1}$.
As a consequence, the slope value of these low-mass galaxies is $\alpha_{\mathrm{sim}} \approx -2.36$ ($N_{\mathrm{gal}} = 43211$).

For the cluster formation efficiency, which equals the bound fraction introduced in \Cref{equ:bound_fraction} and the survival rate of star clusters during the initial few Myr \citep[see][for details on this ``cruel cradle effect'']{kruijssen2012b,kruijssen2012d}, we find an increase with SFR (panel C).
This is expected given the direct relationship between the SFR surface density and cold gas surface density (\textit{c.f.}~\Cref{equ:sfr_density}) and because $f_{\mathrm{bound}}$ positively correlates with $\Sigma_{\mathrm{g}}$ (\textit{c.f.}~\Cref{fig:cluster_bound_fraction}).
Our results are in excellent agreement with literature data.

In the observational studies that we mention in \Cref{sec:results} the cluster formation efficiency is computed by comparing the total mass in young ($\tau_{\mathrm{c}} \leq 10 \, \mathrm{Myr}$) star clusters with the current SFR \citep[e.g.][]{goddard2010a,ryon2014a,hollyhead2016a}.
Broad-band photometry and fits to the inferred spectral energy distribution yield estimates of the star cluster parameters whereas H$\alpha$ or infrared fluxes estimate the star formation rate.

We conclude that this relationship is sensitive to the resolution of the simulation and that taking into account biases in the mass range of the selected galaxy samples is important.

\subsection{Stellar masses}
\label{subsec:stellar_masses}

\begin{figure*}
    \centering
    \includegraphics[width=\textwidth]{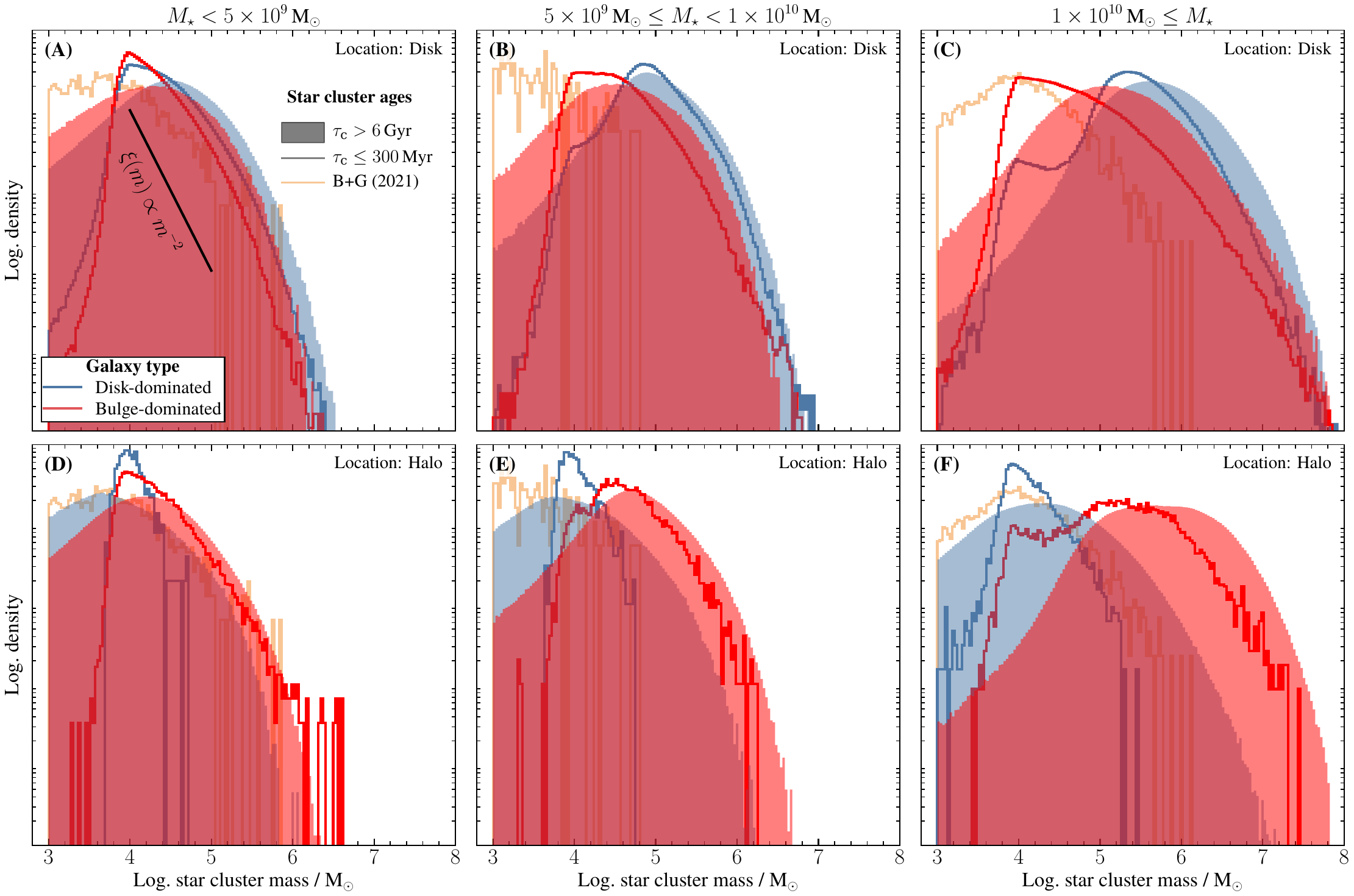}
    \caption{%
        Estimated probability density function values of binned star cluster masses in disk- ($B/T < 0.2$; blue colour) and bulge-dominated ($B/T > 0.7$; red colour) galaxies at $z=0$.
        We track up to \num{2000} star clusters per galaxy, split equally between star clusters located in the disk (\textit{top row}) and halo (\textit{bottom row}).
        Host galaxies are separated into three mass bins.
        Additionally, we separate between young ($\tau_{\mathrm{c}} \leq 300 \, \mathrm{Myr}$; solid lines) and old ($\tau_{\mathrm{c}} > 6 \, \mathrm{Gyr}$; shaded regions) star clusters.
        We sample the cluster initial mass function with an initial mass of $10^{4} \, \mathrm{M}_{\odot}$ and discard star clusters with masses below $10^{3} \, \mathrm{M}_{\odot}$.
        The black solid line in the top left panel presents the asymptotic power-law behaviour towards $m_{\mathrm{c}} = 0$.
        We compare the population of simulated star clusters to the observational resulted of \citet[][orange colour]{brown2021c}.
        Each column shows the same data set as the location of the star clusters (disk versus halo) is not specified by the authors.
    }
    \label{fig:cluster_mass_distribution}
\end{figure*}

We show in \Cref{fig:cluster_mass_distribution} the one-dimensional histogram of star cluster masses at $z=0$.
The galaxies are split into different morphologies and mass bins, and star clusters are separated by their age and location (disk versus halo).
There exist two sharp truncations at mass values of $10^{3}$ and $10^{4} \, \textrm{M}_{\odot}$, which correspond to the cut-off and lowest initial star cluster masses of the simulation.
Both features disappear towards more massive galaxies as they contain more than \num{1000} star clusters in disks, thus, increasing the lowest star cluster mass that we evolve.\footnote{The same trend is apparent for star clusters in bulge-dominated galaxies given their rich merger history compared to their disk-dominated counterparts.}

Young star clusters ($\tau_{\mathrm{c}} \leq 300 \, \textrm{Myr}$) follow approximately a power-law distribution that is truncated towards the high-mass end, according to the initial star cluster mass function.
Higher-mass galaxies can feature more massive star clusters, which is both a statistical effect from sampling the cluster initial mass function as well as higher upper truncation masses from a larger Toomre mass and bound fraction (\Cref{equ:cimf_truncation}).
Apparent differences in the histograms for young star clusters between galaxy morphologies exist mostly because of the technical limitation of only tracking a certain number of star cluster per galaxy.
{\lgalaxies} assumes that star formation only occurs in galaxy disks, and both morphological types are expected to have similar disk properties, leading to the expectation of similar power-law distributions of star cluster masses \citep[see][for details]{henriques2015a,henriques2020a}.
\footnote{There still remains some difference in the metallicities of newly formed star clusters between galaxies of different morphology, as we show in \Cref{subsec:metallicity}.}

Looking at old ($\tau_{\mathrm{c}} > 6 \, \textrm{Gyr}$) star clusters we find broad Gaussian-like distributions.
They broaden and shift towards higher star cluster masses due to a typically higher upper truncation mass of the initial star cluster mass function and the positive correlation between the number of star clusters and a galaxy's mass.
There exist differences between galaxy morphologies, which become most striking in the most massive galaxies.
The masses of gaseous disks are elevated in disk-dominated systems compared to their bulge-dominated counterparts resulting in a higher number of massive star clusters.
This observation is reversed in galaxy halos, which is related to the merger history of the two morphological types:
galaxy mergers, both minor and major, increase the population of star clusters in a galaxy's halo and this effect occurred more often in $z=0$ bulge- than in disk-dominated galaxies.

Finally, we compare our simulated star clusters to observational constraints from \citet{brown2021c} who analysed the structural properties of young star clusters in a set of nearby galaxies from the \textit{LEGUS} programme \citep[][]{calzetti2015a}.
For the comparison, we apply the same galaxy mass and star cluster age constraints as for the simulated data.
For the comparison with our simulated data, we split the literature values into the same galaxy mass bins ($M_{\star} / \textrm{M}_{\odot} < 5 \times 10^{9}$, $5 \times 10^{9} \leq M_{\star} / \textrm{M}_{\odot} < 10^{10}$, and $10^{10} \leq M_{\star} / \textrm{M}_{\odot}$) and applied the same constraints on star cluster ages ($\tau_{\mathrm{c}} \leq 300 \, \mathrm{Myr}$ for young and $\tau_{\mathrm{c}} > 6 \, \mathrm{Gyr}$ for old objects).

Star cluster ages in the analysis from \citet{brown2021c} stem from fitting the spectral energy distribution; see details in \citet[][]{adamo2017a} and \citet[][]{ryon2017a}.\footnote{The data set of \citet{brown2021c} features only a few star clusters older than $6 \, \textrm{Gyr}$. We omit a comparison for old star clusters.}
For dwarf galaxies, we find excellent agreement between the stellar mass distributions of the populations of simulated and observed star clusters down to the sampling limit of $10^{4} \, \textrm{M}_{\odot}$.
In more massive galaxies the comparison becomes more challenging due to a limited number of star cluster that we consider per galaxy.
Nevertheless, we find that our simulated galaxies form more massive star clusters than observed in nearby systems, which could be related to multiple effects.
First, to some degree, the comparison could suffer from poor number statistics from the observations, which does not populate well the high-mass end of star clusters.
Second, our data contain highly star forming galaxies with star formation rates above $10 \, \textrm{M}_{\odot} \, \textrm{yr}^{-1}$ at the high galaxy-mass end, which results in a larger probability to form massive star clusters.
While our assumption for the formation of star clusters, such as the upper truncation mass of the cluster initial mass function, could be inaccurate and lead to elevated star cluster masses, the agreement in the $M_{V}$-SFR relationship that we presented in the previous section would indicate that we sample well the initial star cluster masses.
Finally, the evolution of the young star clusters is sensitive to their initial properties and affected by their direct surroundings.
As we present in the next section, there is a disagreement between the half-mass radii of the simulated and observed star clusters, which could drive some discrepancy between the observed mass functions, given the coupled differential equations of the evolution of a star cluster's mass and radius.

\subsection{Half-mass radii}
\label{subsec:half-mass_radii}

\begin{figure*}
    \centering
    \includegraphics[width=\textwidth]{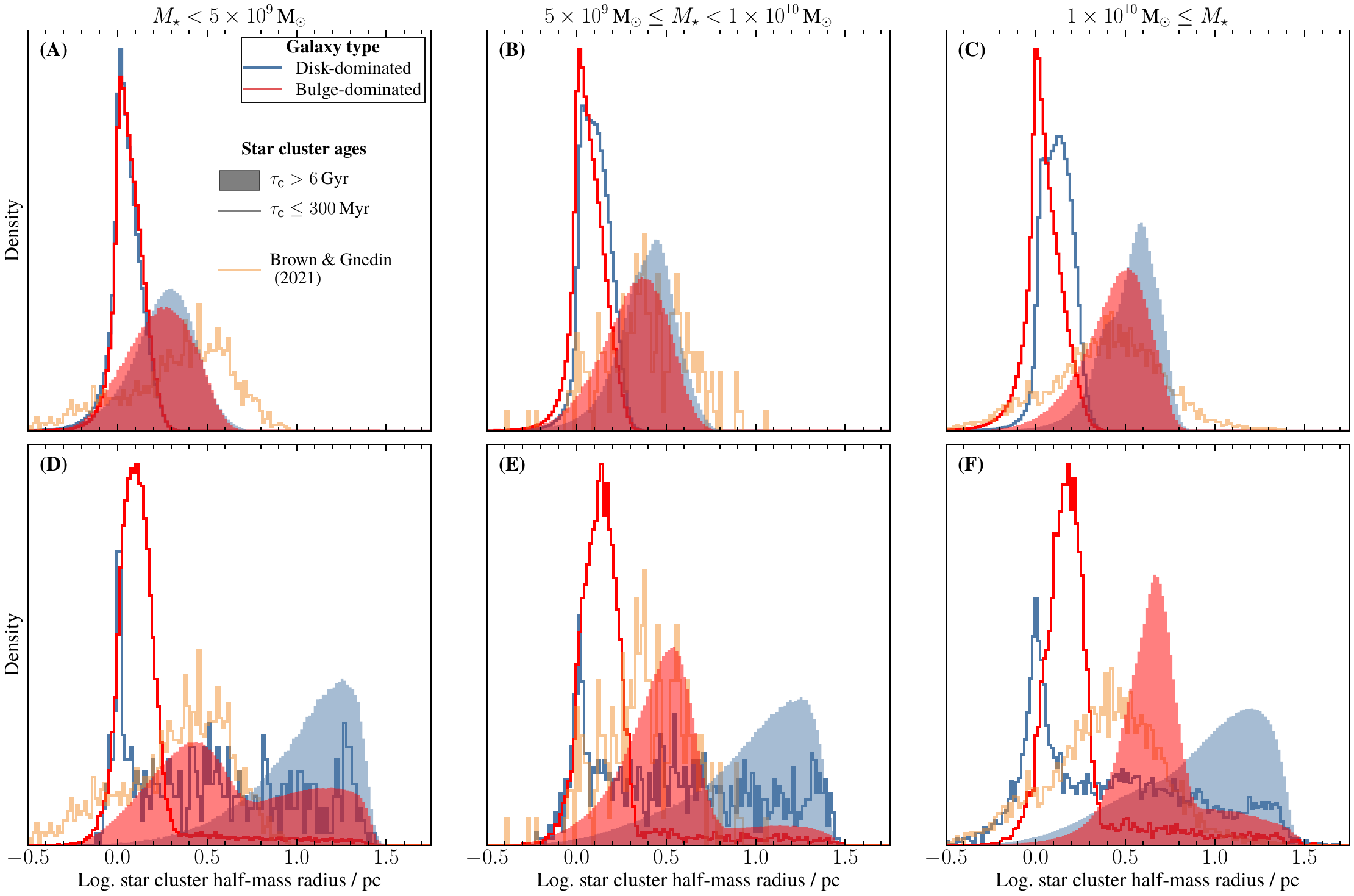}
    \caption{%
        Same as \Cref{fig:cluster_mass_distribution} but for star cluster half-mass radii.
        The double-peaked structure of half-mass radii in a galaxy's halo is a result of the prescription for star cluster re-distribution during galaxy mergers (see \Cref{subsubsec:re-distribution_during_galaxy_mergers} and \Cref{subsec:galactocentric_distances} below).
    }
    \label{fig:cluster_radius_distribution}
\end{figure*}

We present in \Cref{fig:cluster_radius_distribution} the distribution of half-mass radii of star clusters at $z=0$.
The separation into different star cluster ages and galaxy masses follows the previous section.

We find that the half-mass distribution of young star clusters peak around the initial value of $1 \, \textrm{pc}$ and extends in both directions.
Going from low- to high-mass galaxies we find that young star clusters become larger, which is a consequence of the increased average star cluster mass (see \Cref{equ:evolution_half-mass_radius} and the related individual mass-loss terms).
For the same reason, we observe the opposite trend with galaxy morphology between young star clusters in the disk and in the halo (compare to \Cref{fig:cluster_mass_distribution}).\footnote{Young star clusters in a galaxy's halo do not form there. Instead, they formed in a satellite galaxy that was accreted shortly after their formation.}

Older star clusters in galaxy disks show a broader distribution of half-mass radii, which is related to both a broad distribution of their masses and a great variety of their environments, which affect the contributions of tidal shocks to the radial expansion or contraction (\Cref{equ:time_shocks}).
Additionally, the half-mass radii increase, ranging between a few parsecs in dwarf galaxies to ten parsecs in the most massive galaxies.
The half-mass radii of star clusters in disks is limited due to the influence of tidal shocks on the mass: if a star cluster expands significantly, the mass-loss term for tidal shocks increases, thus reducing both the stellar mass and the leading to a contraction of the star cluster.
This coupling then leads to either the destruction of star clusters that experience repeated episodes of radial expansion and contraction, or restrains the half-mass radii to small values.

Older star clusters in a galaxy halo, especially in bulge-dominated galaxies, show a peculiar distribution with a peak around a few parsecs and a separate extended tail towards large half-mass radii.
Both trends are a direct consequence of our assumed prescription for the re-distribution of star clusters during galaxy mergers, which we will detail in the next section.
The peak around a few parsecs comes from major merger events and matches the star cluster distribution in disks, i.e.\ the population of star clusters, which were heated during the galaxy-galaxy interaction.
In contrast, star clusters with large half-mass radii got accreted during minor merger events and are located at great distances from their new host galaxy.
These star clusters do not suffer from tidal shocks.
Instead, the large tidal radius of the star clusters enables two-body relaxation to increase their half-mass size to large values, often exceeding $100 \, \mathrm{pc}$ (see \Cref{equ:evolution_half-mass_radius} and \Cref{equ:evolution_half-mass_radius_zeta_xi}).

We compare our simulated young star clusters to the results of \citet{brown2021c} after applying the same constraints on the star cluster age and galaxy mass.
The simulated star clusters show a much narrower distribution that does not extend to large radii.
Instead, the half-mass radii of observed young star cluster fit well to the distribution of old star clusters in our simulation.
If the data by \citet{brown2021c} present a representative distribution of half-mass radii shortly after cluster birth then they indicate that the radii are potentially constrained by the clusters environment and may evolve quickly \citep[see e.g.][for $N$-body simulations]{banerjee2017a}.
Evidence for the latter comes from hydrodynamical simulations of \citet[][]{lahen2025a} that show a rapid half-mass radius evolution from $\approx 0.1$ to $\approx 1 \, \textrm{pc}$ over $100 \, \textrm{Myr}$.
We explore different prescriptions for initial half-mass radii in \Cref{app:subsec:initial_half-mass_radius} but note that \citet{reina-campos2023b} found that none of their prescriptions, including one that depends on the local environment, can recover the $z = 0$ distribution of the Milky Way.

\subsection{Galactocentric distances}
\label{subsec:galactocentric_distances}

\begin{figure*}
    \centering
    \includegraphics[width=\textwidth]{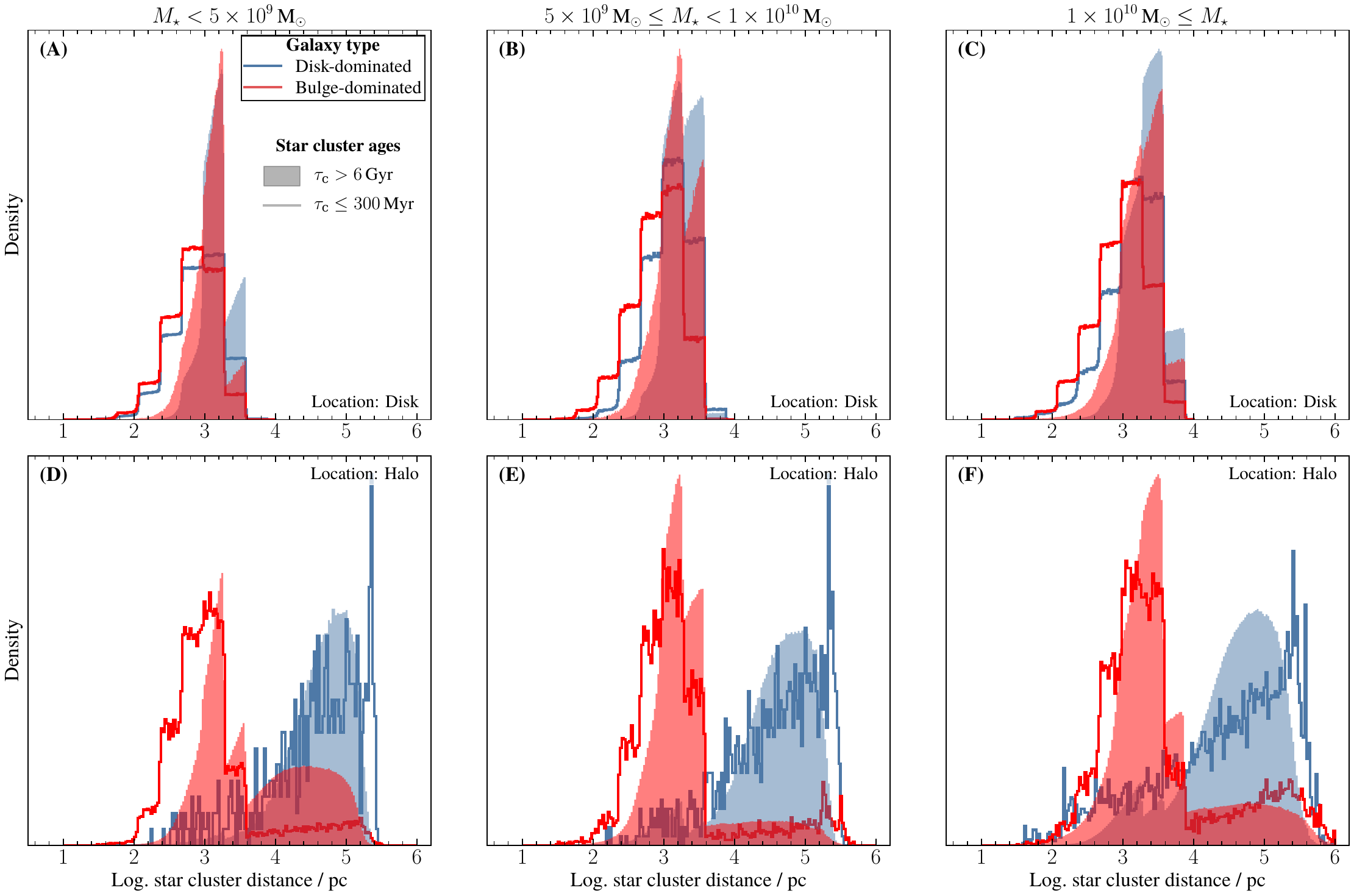}
    \caption{%
        Same as \Cref{fig:cluster_mass_distribution} but for star cluster half-mass radii.
        The double-peaked structure of half-mass radii in a galaxy's halo is a result of the prescription for star cluster re-distribution during galaxy mergers.
    }
    \label{fig:cluster_distance_distribution}
\end{figure*}

We show the distribution of galactocentric distances of star clusters in \Cref{fig:cluster_distance_distribution}.
The separation into different star cluster ages and galaxy masses follows the previous two sections.

Looking first at star clusters within a galaxy disk, we find the `step-like' behaviour of young star clusters, which shows the assumed logarithmically-flat distribution of star cluster in galactocentric distances.
Deviations from this distribution only occur for young star clusters in the central few tens of parsecs, where the local density is highest, the most massive star clusters out to a few hundred parsecs, or old star clusters.
The latter effect is clearly visible by considering only star clusters older than $\tau_{\mathrm{c}} > 6 \, \mathrm{Gyr}$.
Within $\approx 1000 \, \mathrm{pc}$ star clusters efficiently sink towards the center and either merge within the central few parsecs to form a nuclear star cluster, whose formation we do not consider here, or get tidally destroyed due to the increasing gas density and fraction of giant molecular clouds.

For star clusters in the halo we find two separate distributions.
First, we find the same `step-like' behaviour as for star clusters in the disk, which come from major mergers: as mentioned in \Cref{subsubsec:re-distribution_during_galaxy_mergers} we assume that star clusters located in the disk of the more massive galaxy get heated to the halo but retain their galactocentric distance during major galaxy mergers.
This effect, therefore, is mostly seen in bulge-dominated galaxies given that most $z=0$ disk-dominated galaxies never experienced a major merger.
In contrast, we assume that the less massive galaxy gets tidally disrupted in (minor and major) galaxy mergers, resulting in a large range of galactocentric distances.

Given our assumptions on how star clusters accrete, we find that the \textit{in-situ} star cluster population, that is present in the disk, dominates in number over the \textit{ex-situ} star cluster population within a few to $\approx 10 \, \mathrm{kpc}$, depending on the host galaxy mass and type.
These values are in rough agreement with the values presented in \citet{keller2020b} for the {\emosaics} simulation, however, the shape of the distribution of star cluster galactocentric distances is clearly different.
A more realistic `smooth' star cluster distribution as a function of distance that originates from major mergers requires a more accurate treatment for the more massive galaxy during galaxy mergers within {\lgalaxies}, which is beyond the scope of this work.

\subsection{Metallicity distributions}
\label{subsec:metallicity}

\subsubsection{Mean cluster metallicity}
\label{subsubsec:mean_cluster_metallicity}

\begin{figure*}
    \centering
    \includegraphics[width=\textwidth]{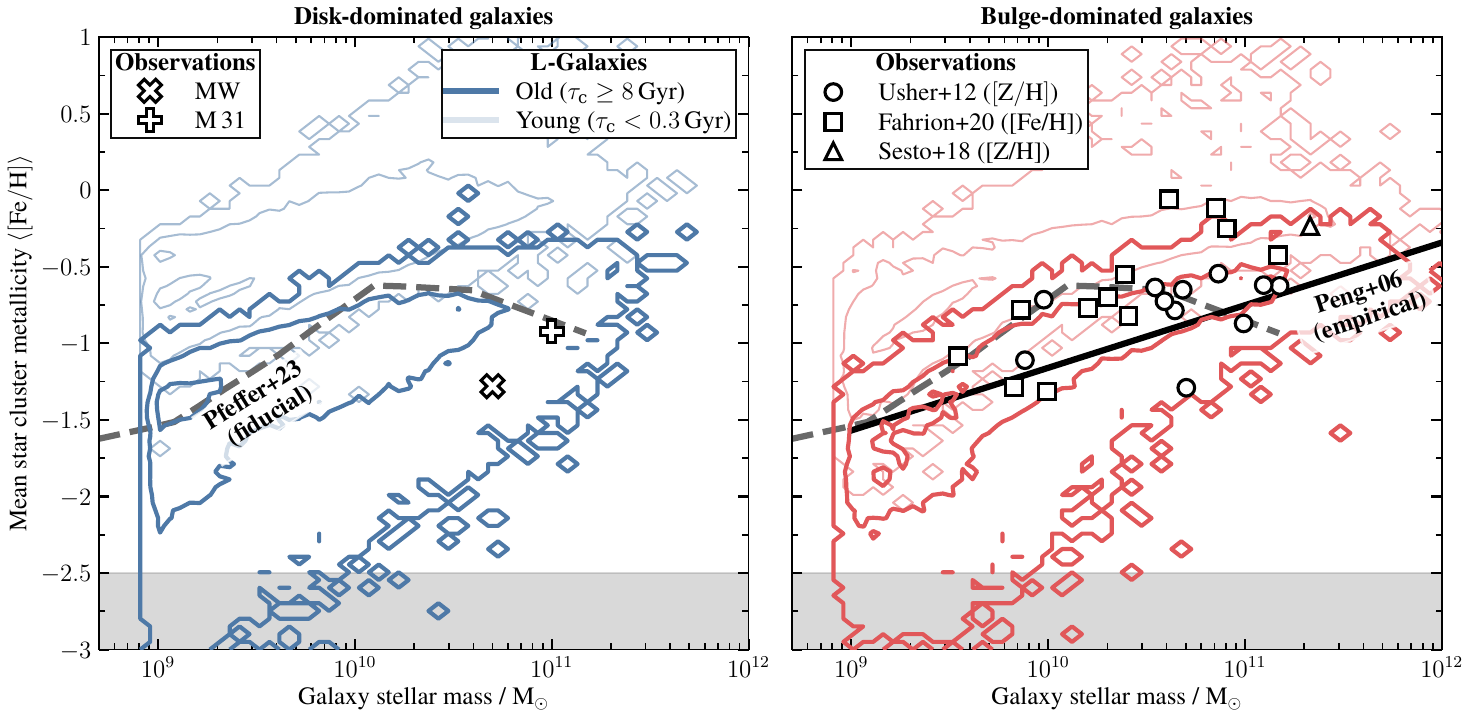}
    \caption{%
        Mean star cluster metallicity per galaxy versus host galaxy stellar mass, separated into disk- (\textit{left panel}) and bulge-dominated (\textit{right panel}) galaxies.
        Strong / faint contours give the 1-, 2-, and 3-$\sigma$ distribution for old / young star clusters.
        The gray dashed line gives the fiducial model of \citet{pfeffer2023b}, which assumes, similar to our model, an environmentally-dependent prescription for the upper truncation mass of the cluster initial mass function and the cluster formation efficiency.
        The black solid line gives the empirical relationship for ellipticals in the nearby Virgo galaxy cluster \citep{peng2006a}.
        Data for the Milky Way and M{\,}31 stem from a self-compiled data table that will be presented in future work.
        Other data points come from \citet{usher2012a}, \citet{sesto2018a}, and \citet{fahrion2020b}.
        The gray-shaded area marks the ``lower-limit floor'' at $\langle [\mathrm{Fe} / \mathrm{H}] \rangle = \num{-2.5}$ of observed star cluster metallicities in other galaxies \citep[e.g.][]{beasley2019a}.
    }
    \label{fig:cluster_mean_metallicity_galaxy_mass}
\end{figure*}

We show in \Cref{fig:cluster_mean_metallicity_galaxy_mass} the mean metallicity of a galaxy's star cluster population, as traced by their iron-abundances, versus galaxy stellar mass.
The iron abundance values are calculated by the galactic chemical evolution model introduced into \textit{L-Galaxies} by \citet{yates2013a} that takes into account contributions from stellar winds and (type-Ia and -II) supernovae with different delay-time-distributions.
We split the galaxies by their morphological type (disk-dominated versus elliptical; see above) and separate the star cluster population by their age (young versus old; see above).

Irrespective of the star clusters age and galaxy morphology we find that the mean metallicity increases as a function of galaxy mass, similar to the mass-metallicity relationship for galaxies \citep[as traced via oxygen-abundances; e.g.][]{tremonti2004a,kewley2008a,torrey2019a,sanders2021a}.
Despite some overlap, younger star clusters have higher metallicity than their older counterparts at fixed stellar mass for both galaxy types.
This difference appears to be more significant for disk- than for bulge-dominated galaxies, which is related to the origin of the cold gas that forms stars:
in disk-dominated systems stars are predominantly formed from gas that has been continuously enriched by the above-mentioned stellar winds and supernovae channels resulting in metal-rich young star clusters.
In bulge-dominated galaxies that form stars (and star clusters) a significant fraction of the cold gas comes from accreted lower-mass systems that contain relatively metal-poor gas.

Towards the low-mass end, we find that a significant fraction of galaxies ($32.1 \, \%$ and $53.0 \, \%$ for disk- and bulge-dominated galaxies below $10^{10} \, \textrm{M}_{\odot}$, respectively) have at least one star cluster with a metallicity value $\langle [\textrm{Fe}/\textrm{H}] \rangle \leq -2.5$. In contrast, no disk- and only $0.003 \, \%$ of bulge-dominated galaxies have a mean star cluster metallicity below $-2.5$.
This threshold value of $-2.5$ is often used to indicate a ``metallicity floor'' due to an apparent lack of globular clusters in nearby galaxies below this value \citep{beasley2019a}.\footnote{However, the detection of a low-metallicity stellar stream of a former massive star cluster in the Milky Way halo \citep{martin2022a} and the detection of a massive star cluster with $[\mathrm{Fe} / \mathrm{H}] \approx -2.9$ in M{\,31} \citep{larsen2020a} challenge this notion.}
The contours of our simulated data drop down to extremely low metallicities at the mass-resolution limit.
When running the model on the Millennium-II data set, the star cluster distributions follow more closely the scaling relation by \citet{peng2006a}, indicating that resolution effects play a role at the low-mass end.

We compare our results to observational data.
For disk-dominated galaxies, we consider the Milky Way and M{\,}31 as they have the most robust and quantitative measurements of globular cluster metallicities.
To obtain the mean metallicity values, we collected the cluster information from a diverse set of literature that we present in a future paper \citep[another table is presented in][]{pace2025a}.
Here we find that our simulated disk-dominated galaxies have, on average, about $0.3 \, \textrm{dex}$ higher median metallicity than the observed values in the Milky Way, which might be related to a (1) a lack of accreted dwarf galaxies that would contribute mainly low-metallicity star clusters, (2) a too aggressive prescription for the destruction of star clusters with low metallicity or old age, or (3) a computational limitation in that we only store the properties of the $1000$ most massive star clusters in a disk.
\citet{pfeffer2023b} find a better agreement with observational constraints when limiting their sample to a galactocentric distance of $20 \, \textrm{kpc}$, thus, excluding preferentially low-metallicity star clusters. We explore this issue again in a future paper.

For bulge-dominated galaxies we collect data from \citet{usher2012a,sesto2018a,fahrion2020b} and take the empirical scaling relation from \citet{peng2006a}.
Our simulated star cluster populations show excellent agreement with the observations across the whole mass scale.
A couple of literature data points lie outside the \num{3}-$\sigma$ contours, which could be related to either poor number statistics of the or a bias in the ages of the star clusters in the observations with respect to our simulated galaxies.

\subsubsection{Bimodality}
\label{subsubsec:bi-model_metallicity_distributions}

An extensive set of literature work argues that the star cluster population of many, perhaps all, galaxies shows a bi- or multi-modal colour or metallicity distribution \citep[e.g.][]{cohen1998b,kundu2001a,beasley2008a,brodie2012a,blom2012c,usher2012a,usher2013a,escudero2015a,caldwell2016a,bassino2017a,villaume2019a,fahrion2020b,hixenbaugh2022a,lomeli-nunez2024a}.
An often referred to explanation is that the bi-modality is a result of different origins of clusters with the bluer (more metal-poor) population having an \textit{ex-situ} origin \citep[e.g.][]{strader2005a,brodie2006a,katz2013a,tonini2013a}.
Numerical work has argued that the bi- or multi-modality could be a result of either different epochs of star cluster formation, different cluster origins, as suggested by the above literature, or details related to cluster formation and destruction, or some combination thereof \citep[see e.g.][]{kruijssen2015b,choksi2019b,chen2025a}.
However, as pointed out by the observational work of \citet{pastorello2015a} and the numerical findings by \citet{pfeffer2023b} the bimodality may only be present in a minority ($\lesssim 50 \, \%$) of systems.

To determine the bimodality in our simulation we use an Bayesian Gaussian Mixture Model approach with Dirichlet initial conditions for all galaxies than contain more than 30 clusters.
Following \citet{muratov2010a} and \citet{pfeffer2023b} we first determine
\begin{equation}
    -2 \ln \lambda = -2 \ln \bigg[ \frac{\max (\mathcal{L}_{1})}{\max (\mathcal{L}_{2})} \bigg] \;,
    \label{equ:bgmm_lambda}
\end{equation}
where $\max (\mathcal{L}_{j})$ is the maximum value of the likelihood function evaluated over the metallicity distribution when considering $j$ number of Gaussians (either one or two).
Afterwards, we perform 100 bootstrap iterations to evaluate the probability that the solution is bi-modal with a probability threshold of $90 \, \%$, as chosen in \citet{pfeffer2023b}.\footnote{We confirmed that the random sampling of initial values during bootstrapping affects the bi-modal fractions at most by 0.05.}
Finally, we calculate the weighted distance between the two Gaussians,
\begin{equation}
    D_{\mathrm{G}} = \frac{|\mu_{1} - \mu_{2}|}{\sqrt{\sigma_{1}^{2} + \sigma_{2}^{2}}} \overset{!}{\geq} \sqrt{2} \;,
    \label{equ:bgmm_distance}
\end{equation}
where $\mu_{j}$ and $\sigma_{j}$ are the mean and standard deviations of Gaussian $j$, respectively.
If $D_{\mathrm{G}} < \sqrt{2}$, we classify the distribution as uni-modal, as adopted in \citet{muratov2010a} and \citet{pfeffer2023b}.\footnote{Note that we do not utilise a cut on the kurtosis of the Gaussians as we find in some cases that, if there is one dominant and another sub-dominant star cluster population in a galaxy, the kurtosis is greater than zero, indicating a ``peaked'' distribution, thus, rejecting the hypothesis of a bimodal distribution.}

\begin{figure}
    \centering
    \includegraphics[width=\columnwidth]{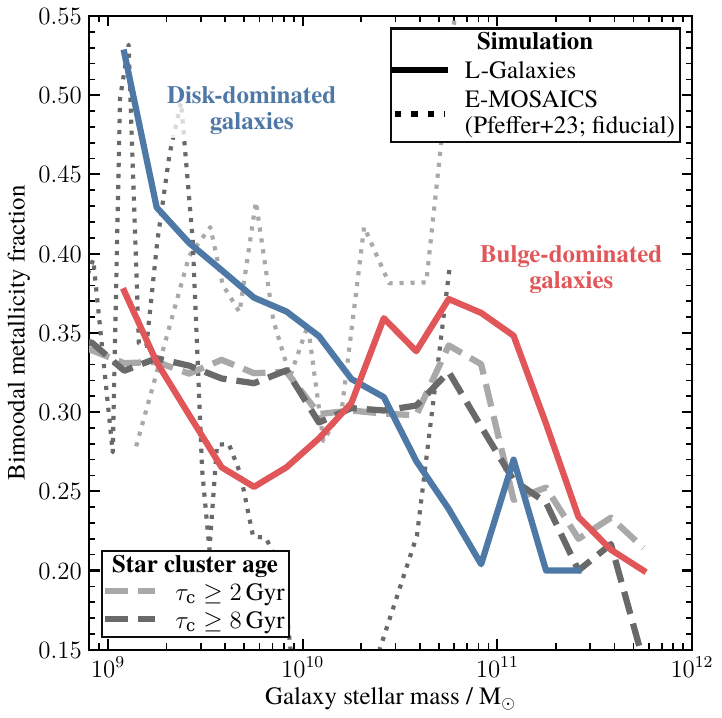}
    \caption{%
        Fraction of galaxies exhibiting a bimodality in the metallicity distribution of their star cluster population versus stellar mass for disk- (blue) and bulge-dominated (red) galaxies.
        We show the bimodality when restricting the cluster population to ages $\tau_{\mathrm{c}} \geq 2 \, \textrm{Gyr}$ (light gray) and $\tau_{\mathrm{c}} \geq 8 \, \textrm{Gyr}$ (dark gray) in order to compare the results to the ones for the {\emosaics} simulation from \citet[][dotted lines]{pfeffer2023b}.
    }
    \label{fig:cluster_bimodal_fraction_galaxy_mass}
\end{figure}

We show the bimodality of the star cluster distribution as a function of galaxy stellar mass in \Cref{fig:cluster_bimodal_fraction_galaxy_mass} where we split all galaxies into disk- and bulge-dominated.
Irrespective of galaxy morphology we find that the bimodal fraction (i.e.\ the fraction of galaxies that exhibits a bimodal metallicity distribution) is constrained to values ranging between $\approx 20$ and $\approx 50 \, \%$.

In disk-dominated galaxies, the bimodality of the star cluster population decreases with increasing galaxy stellar mass, which is attributed to both a continuous formation of star clusters of increasing metallicity, which does not show an intrinsic bimodal distribution at $z=0$, and an increasing number of minor merger events, which bring in additional star cluster populations and diffuse any potential bimodality that would exist in the data.
The latter argument can be used to explain why the bimodality decreases for the most massive bulge-dominated galaxies, where these systems experienced, on average, more minor and major mergers than their disk-dominated counterparts.
As a result the overall metallicity distribution at $z = 0$ extends over a large range and individual sub-structures or features become diffused.
This effect can be seen in observations as well, such as massive galaxies like M{\,}87 \citep[see Fig.\ \num{20} in][]{cohen1998b} or massive galaxy clusters \citep[e.g.][]{harris2017a}.

For bulge-dominated systems specifically, we find a dip in the bimodal fraction around $8 \times 10^{9} \, \textrm{M}_{\odot}$ by up to ten percent.
This decrease is related to an increased number of accreted star clusters in comparison to \textit{in-situ} star clusters, and is not a consequence of a sudden change in the number or type of galaxy mergers.
Towards higher galaxy masses, the \textit{in-situ} fraction of star clusters increase again and result in an increase of the bimodality, continuing to a few times $10^{10} \, \textrm{M}_{\odot}$ after which the number of accreted star clusters dominate again.

We compare our results to the {\emosaics} simulation where \citet{pfeffer2023b} performed part of the metallicity analysis.
Note that, for the comparison, we split the star clusters into two age bins, $\tau_{\mathrm{c}} \geq 2 \, \textrm{Gyr}$ and $\tau_{\mathrm{c}} \geq 8 \, \textrm{Gyr}$, to match the selection criteria by \citet{pfeffer2023b}.
When looking at the bimodal fractions with respect to an age cut in the cluster populations, we find good agreement between the two models, although our simulation provides a higher number of galaxies, thus avoiding statistical fluctuations.
While there appears no significant difference between star cluster populations when imposing the age constraints, there appears to be a slight difference compared to the general star cluster populations in dwarf galaxies.
This result shows that the bimodal classification in dwarf galaxies can be significantly influenced by newly forming star clusters whereas the total number of star clusters in more massive galaxies far outweighs any newly formed population.

The bimodality weakly depends on the location within a galaxy.
Compared to the average bimodal fraction of all star clusters we find similarity for the inner-most galaxy regions, an increase at intermediate distances, and statistical fluctuations for the outer regions, all normalised to the disk's scale-length and half-mass radius for disk- and bulge-dominated, respectively.
This trend is a consequence of the location of accreted star clusters as the central regions are dominated by \textit{in-situ} star clusters and the outer regions by \textit{ex-situ} ones.
The overall trend of the bimodal fraction with galaxy stellar mass remains roughly unchanged.

Our results indicate a lower bimodality when compared to observations and the general notion that a bimodal distribution appears frequently in galaxies.
To test the origin of this notion and the difference we compile a list of galaxies that were classified as uni- or bimodal from the above-mentioned literature and bin the data by galaxy mass.
Overall we find bimodal fractions ranging between $50$ and $100 \, \%$, however, galaxy number statistics are low.
It appears likely that the discrepancy between these results and our simulations are due to differing methodologies and inhomogeneities when analysing data.
For example, several studies consider a bimodality in the colour-distribution whereas others use iron abundances.
Furthermore, taking the metallicity values of globular cluster candidates from \citet{beasley2008a} for NGC{\,}5128 and applying our methodology results in a classification of uni-modal (because the weighted distance $D_{\mathrm{G}}$ is smaller than $\sqrt{2}$) whereas the authors classify the distribution as multi-modal.
Unfortunately, the metallicities of many star clusters in galaxies are not publicly available and do not allow us to test the effect of different methodologies further.

\subsection{Caveats}
\label{subsec:caveats}

Our semi-analytical approach to model star cluster populations comes with several caveats.
Here we summarise a few of the most important issues.

\subsubsection{Axisymmetric structures of galaxies}
\label{subsubsec:axisymmetric_structures_of_galaxies}

Several effects of non-axisymmetric features can impact star clusters, some of which are discussed below.
A thorough review is presented in \citet[][]{renaud2018b}.

Some basic assumptions of {\lgalaxies} potentially break down as the gas fraction of galaxies increases, resulting in a ``clumpier'' structure \citep[e.g.][]{conselice2004a,foerster-schreiber2011a,shibuya2016a}.
This raises some doubt about whether our model captures the properties of star clusters at redshifts of about one to two where clumpy galaxies start to make up a dominant fraction of the whole galaxy population \citep[e.g.][]{sattari2023a,huertas-company2024a} and whether the $z = 0$ star cluster population evolved in a similar fashion compared to observed star clusters.
However, as argued by \citet{ono2025a} the basic principle of a disk-dominated formation scenario for galaxies may be a reasonable assumption for redshifts of $z \gtrsim 9$.

A related issue is that {\lgalaxies} does not consider non-axisymmetric components like bars.
Molecular gas can gather at tips of the bars due to orbital crowding \citep[][]{kenney1991b} leading to collisions of GMCs, triggering the formation of stars and star clusters \citep[e.g.][]{davies2012a,fukui2014a,ramirez-alegria2014a}.
A bar can influence the dynamical evolution of star clusters as well.
For instance, both \citet{bajkova2023a} and \citet{dillamore2024b} argue that the Galactic bar directly influences the orbit of a number of Milky Way globular clusters, possibly supporting accelerating their radial migration to the Galactic Center.
While halo star clusters are most likely not significantly affected by the lacking implementation of bars, objects in the galactic disk will likely be influenced, thus potentially changing the expected galactocentric distribution discussed in \Cref{subsec:galactocentric_distances}.

Finally, {\lgalaxies} does not consider the existence of spiral arms.
As argued by \citet{saha2010b}, (transient) spirals and bars can introduce additional energy sources for tidal heating that would increase the orthogonal velocity dispersion of star clusters compared to the orientation of the galactic disk.
This effect could thus contribute star clusters to the halo of a galaxy without invoking galaxy-galaxy interactions, as indicated by the presence of metal-rich open clusters in the Milky Way's halo \citep[see][for examples]{paunzen2006d,meibom2009b,brogaard2012b,heiter2014a,oenehag2014a,straizys2014a,gustafsson2016a,hunt2023b}.

\subsubsection{Impact of stellar-mass black holes}
\label{subsubsec:impact_of_stellar-mass_black_holes}

Over time, stellar mass black holes segregate towards the cluster's centre and build up a dense core, injecting energy into the stellar system of the star cluster \citep[e.g.][]{merritt2004d,mackey2008b}, possibly leading to the star clusters dissolution \citep[e.g.][]{giersz2019a}.
This can result in some star clusters having relatively large half-mass radii, such as Palomar{\,}5 \citep[][]{gieles2021a}, rivalling the most extended objects we find in our simulation.
However, note that we only consider star-star interactions in the formalism of \Cref{equ:time_relaxation}, thus neglecting the effect of dynamical heating due to black holes.
One consequence of adding feedback from black holes is a metallicity-dependent expansion rate that was already explored in the literature \citep[e.g.][]{downing2012a,mapelli2013b,banerjee2017b,chattopadhyay2022a,rostami-shirazi2024d}.
This may result in a change in half-mass radii of young star clusters between different galaxy morphologies that we do not detect in \Cref{fig:cluster_radius_distribution}.
We aim to implement this feedback channel in future versions.

\subsubsection{Galaxy-galaxy interactions}
\label{subsubsec:galaxy-galaxy_interactions}

Interactions between galaxies result in collisions between gas clouds and can cause efficient star formation outside of galaxy disks, often including the formation of star clusters \citep[e.g.][]{fellhauer2005a,annibali2011b,maji2017c,randriamanakoto2019a,rodruck2023a}.
Such star clusters, especially during the first passage of the accreted galaxy, can survive for significant time and could contribute to the $z=0$ globular cluster population in the halo \citep[e.g.][]{li2022a}.
\citet{keller2020b} find that around $20 \, \%$ of globular clusters form during galaxy-galaxy merger events.
If true, this would indicate that our model approach does not explain the origin of a significant fraction of globular clusters at $z=0$.
Nevertheless, as argued by the authors, repositioning of globular clusters from the dense inner-galactic regions into a galaxy's halo is important for cluster survival, which matches our results.

\section{Conclusions}
\label{sec:conclusions}

We introduced a modified version of the semi-analytical galaxy formation model ``{\lgalaxies}'' \citep[][]{henriques2020a,yates2021a} that accounts for the formation of massive ($m_{\mathrm{c}} \geq 10^{4} \, \textrm{M}_{\odot}$) star clusters.
This implementation relies on galaxy constituents to derive the bound fraction of star formation and the total star cluster mass via {\lgalaxies}' prescription of star formation within galaxy disks.
Star cluster masses are random realisations of an environmentally-dependent cluster initial mass function, that is assumed to be a truncated power-law function, and are assigned initial half-mass radii, metallicities, and galactocentric distances.
We evolve the properties of up to 2000 individual star clusters per galaxy taking into account the effects of stellar evolution, two-body relaxation, tidal shocks, dynamical friction, and a redistribution during galaxy mergers.

Running the simulation on output merger trees from the Millennium \citep[][]{springel2005c} and Millennium-II \citep[][]{boylan-kolchin2009a} simulations yields the following results.
\begin{itemize}
    \item The most massive and young ($\tau_{\mathrm{c}} < 0.3 \, \textrm{Gyr}$) star clusters in disk-dominated galaxies follow the observed empirical relationship between their absolute $V$-band magnitude of the total host galaxies star formation rate. There exist secondary dependencies on the host galaxy's stellar mass and cluster formation efficiency; the convolved bound fraction of star formation and initial star cluster survival rate.
    \item The assumption of the relationship between the sound speed of cold gas in the interstellar medium and the surface star formation rate directly influences the properties of young star clusters. For example, assuming that the turbulence in the interstellar medium is mainly related to gravity results in a slope value of the relationship between the $V$-band magnitude and the star formation rate being too shallow.
    \item The star cluster mass function for young clusters exhibits a profile similar to the observational results of nearby disk-dominated galaxies \citep[][]{brown2021c}. More massive galaxies host both a larger number of star clusters and more massive ones.
    \item The half-mass radii of star clusters are limited to about ten parsecs in galaxy disk. In galaxy halos, depending on the galactocentric distance, star clusters can expand to half-mass radii greater than $100 \, \mathrm{pc}$. Our data cannot reproduce well the observational constraints of \citet{brown2021c}. While specific choices of the distribution of initial half-mass radii can sometimes match the observations in specific galaxy mass ranges, no prescription can match the observational constraints in all galaxy types at all galaxy masses. Reproducing these observations remains challenging for current simulations.
    \item The mass, half-mass radius, and galactocentric distance functions of old ($\tau_{\mathrm{c}} \geq 6 \, \textrm{Gyr}$) star clusters display complex shapes, which results from different star cluster origins (\textit{in-situ}, accreted, heated during galaxy mergers), and environmentally-dependent prescriptions that impact star cluster evolution. Galaxy-galaxy interactions and mergers play a vital role in shaping the properties of the $z = 0$ star cluster population.
    \item Our model is in excellent agreement with observations of the correlation between a bulge-dominated galaxy's mean star cluster metallicity, as traced by iron abundance, and its host galaxy stellar mass over four dex. This result corroborates the importance of taking into account metal enrichment of the circum-galactic medium from supernovae, as introduced in \citet{yates2021a}.
    \item Our results predict that the presence of a bimodality in the metallicity distribution of star clusters decreases from $\approx 50$ to $\approx 20 \, \%$ with galaxy mass at $z=0$. We do not find a bimodality significantly above $50 \, \%$ for any galaxy morphology at any mass scale. Both different methodological approaches and the inaccessibility of a statistically significant data set of star cluster populations do not allow for a clean comparison between our simulated galaxies and observations.
\end{itemize}
Our simulation offers a computationally efficient and flexible approach to probe different physical effects that influence the assembly history of star clusters across diverse galaxy populations in mass, type, and evolution over cosmic time.
In future work we plan to look at additional aspects of the model, such as the star cluster properties of Milky Way analogues or their evolution with redshift, and to consider the formation of nuclear clusters as well as their interactions with (massive) black holes.

\section*{Data availability}
The code and full data output will be publicised with future papers of the series.
Early access may be granted upon request.

\begin{acknowledgements}
NH thanks F.~Belfiore, I.~{Cabrera-Ziri}, M.~Gieles, J.~Greene, J.~Lee, M.~Mapelli, D.~Maschmann, T.~Naab, M.~{Reina-Campos}, A.~Seth, V.~Springel, and S. Torniamenti for useful discussions.
NH was a Fellow of the International Max Planck Research School for Astronomy and Cosmic Physics at the University of Heidelberg (IMPRS-HD) and acknowledges their support.
NH acknowledges funding through \href{https://www.lacegal.com/}{LACEGAL}, a Latinamerican Chinese European Galaxy Formation Network. This project has received funding from the European Union’s HORIZON-MSCA-2021-SE-01 Research and Innovation programme under the Marie Sklodowska-Curie grant agreement number \num{101086388}.
NH acknowledges funding from the \textit{Centre national d'{\'{e}}tudes spatiales} (CNES) via their postdoctoral fellowship programme.
S.B.\ acknowledges support from the Spanish Ministerio de Ciencia e Innovación through project PID2021-124243NB-C21.
DHC acknowledges support from the Basque Government through the Programa Predoctoral de Formaci{\'{o}}n de Personal Investigador No Doctor del departamento de Ciencia, Universidades e Innovaci{\'{o}}n del Gobierno Vasco.
DS acknowledges the support by the Tsinghua Shui Mu Scholarship, the funding of the National Key R\&D Program of China (grant No.\ 2023YFA1605600), the science research grants from the China Manned Space Project with No.\ CMS-CSST2021-A05, and the Tsinghua University Initiative Scientific Research Program (No.\ 0223080023).
MCA acknowledges support from Fondecyt Iniciaci{\'{o}}n \#11240540.
MP acknowledges funding by the European Union (ERC, Unleash-TDEs, project number 101163093). Views and opinions expressed are however those of the author(s) only and do not necessarily reflect those of the European Union or the European Research Council. Neither the European Union nor the granting authority can be held responsible for them.
\end{acknowledgements}

\bibliographystyle{aa}
\bibliography{./references.bib}

\appendix

\section{Model variations}
\label{app:sec:model_variations}

We present here additional variations of model parameters that influence the assembly history of $z = 0$ star cluster populations.

\subsection{Velocity dispersion of the cold gas}
\label{subsec:velocity_dispersion_of_the_cold_gas}

One of the most crucial parameters for modelling the Toomre parameter is the velocity dispersion of the cold gas.
As discussed in e.g.\ \citet{lehnert2009a} and \citet{zhou2017a} there exists substantial scatter in the relationship between the velocity dispersion of the cold gas and the star formation rate surface density (\textit{c.f.} \Cref{equ:speed_of_sound_cold_gas}).
In our fiducial model we adopted $\alpha_{\mathrm{cold}}^{\mathrm{fid}} = 5 \, \textrm{km} \, \textrm{s}^{-1}$, $\beta_{\mathrm{cold}}^{\mathrm{fid}} = 20 \, \textrm{km} \, \textrm{s}^{-1}$, and $\gamma_{\mathrm{cold}}^{\mathrm{fid}} = 1 / 3$ for the offset, slope, and exponent, respectively.

Here we explore a wider parameter range by testing a first version where we increase the slope value to $\beta_{\mathrm{cold}}^{\mathrm{var}} = 100 \, \textrm{km} \, \textrm{s}^{-1}$ and a second version where we set the exponent to $\gamma_{\mathrm{cold}}^{\mathrm{var}} = 1 / 2$.
Additionally, we test a separate prescription based on the Jeans mass \citep[see e.g.][]{elmegreen2007b} where
\begin{equation}
    \sigma_{\mathrm{g} ,\, j} \sim M_{\mathrm{J}}^{1/4} \mathrm{G}^{1/2} \Sigma_{\mathrm{g} ,\, j}^{1/4} = 4.4 \, \textrm{pc} \, \textrm{Myr}^{-1} \; \Sigma_{\mathrm{SFR} ,\, j}^{0.18} \;,
    \label{equ:cold_gas_velocity_dispersion_jeans}
\end{equation}
with $M_{\mathrm{J}}$ as the Jeans mass, which we assume to be $10^{9} \, \textrm{M}_{\odot}$ for the equality, and converted from the cold gas surface density to the star formation rate density using Equation~7 from \citet{kennicutt1998a} with $\Sigma_{\mathrm{SFR}}$ in units of $10^{6} \, \textrm{M}_{\odot} \textrm{pc}^{-2} \textrm{Myr}^{-1}$.
In addition, we add a velocity floor of $5 \, \textrm{km} \textrm{s}^{-1}$, the same value that we used in \Cref{equ:speed_of_sound_cold_gas}.

To probe the effect of this relationship on the properties of newly formed star clusters, we fit linear relationships to the resulting $M_{V}$-SFR parameter space, as presented in \Cref{subsec:mv-sfr_relationship}, and present the slope values in \Cref{tab:appendix_slope_values}.

We find almost no difference in the results between the fiducial model and the model where $\gamma_{\mathrm{cold}} = 1 / 2$, indicating that the velocity dispersion of the cold gas has a subdominant effect on the effective Toomre stability parameter in comparison to the velocity dispersion of stars in the disk.
For the case where we increase the importance of star formation by increasing the slope parameter $\beta_{\mathrm{cold}}$ we find a slightly steeper slope than in the fiducial model.
This result highlights that the assumption on the relative importance of the star formation surface densities become increasingly important with lower-mass galaxies.
Here the Toomre stability parameter of the cold gas disk gains more relative weight compared to the one of the stellar disk.
Since $Q_{\mathrm{D} ,\, \mathrm{g}} \propto c_{\mathrm{s} ,\, \mathrm{cold}}$, the Toomre parameter of the gas increases/decreases, which then also results in an increase/decrease of $Q_{\mathrm{eff}}$, albeit not as strong as $Q_{\mathrm{D} ,\, \mathrm{g}}$ due to taking a weighted average with $Q_{\mathrm{D} ,\, \mathrm{s}}$.
This increase/decrease has a direct consequence on the star cluster masses because the upper truncation mass in \Cref{equ:cimf_env,equ:cimf_truncation} scales as $m_{\mathrm{cl} ,\, \mathrm{max}} \propto Q_{\mathrm{eff}}^{4}$, resulting in an increased/decreased probability to randomly sample massive clusters.
The increase in stability is also expected given the mass dependence of the $M_{V}$-SFR relation that we present in \Cref{fig:cluster_mv_sfr}, which comes from the star formation rate versus galaxy mass relationship.
For the same reason, we find a stark decline in the steepness of the relationship when adopting \Cref{equ:cold_gas_velocity_dispersion_jeans}.

Our simulations indicate that the slope of the relationship is steeper for more massive galaxies irrespective of the assumed cold gas velocity dispersion versus surface star formation rate relationship.
\begin{table}
    \caption{%
        Slope values of the $M_{V}$-SFR relationship when adjusting parameters for the $c_{\mathrm{s} ,\, \mathrm{cold}}$-$\Sigma_{\mathrm{SFR}}$ relationship.
    }
    \begin{center}
        \begin{threeparttable}
            \begin{tabular}{lll}
                \toprule
                \multicolumn{1}{c}{Model} & \multicolumn{2}{c}{Slope values} \\
                \cmidrule(lr){2-3}
                 & \multicolumn{1}{c}{high-mass} & \multicolumn{1}{c}{low-mass} \\
                \midrule
                {Fiducial} & $-3.16^{+0.01}_{-0.01}$ & $-2.36^{+0.01}_{-0.01}$ \\ \addlinespace
                {$\beta_{\mathrm{cold}}^{\mathrm{var}} = 100 \, \textrm{km} \textrm{s}^{-1}$} & $-3.15^{+0.01}_{-0.01}$ & $-2.44^{+0.01}_{-0.01}$ \\ \addlinespace
                {$\gamma_{\mathrm{cold}}^{\mathrm{var}} = 1 / 2$} & $-3.16^{+0.01}_{-0.01}$ & $-2.38^{+0.01}_{-0.01}$ \\ \addlinespace
                {$c_{\mathrm{s} ,\, \mathrm{cold}} \propto (M_{\mathrm{J}} \Sigma_{\mathrm{g}})^{1/4}$} & $-2.27^{+0.01}_{-0.01}$ & $-1.50^{+0.01}_{-0.01}$ \\ \addlinespace
            \bottomrule
            \end{tabular}
        \end{threeparttable}
    \end{center}
    \label{tab:appendix_slope_values}
\end{table}

\subsection{Initial half-mass radius}
\label{app:subsec:initial_half-mass_radius}

The distribution of initial half-mass radii remains an unsolved problem.
Measurements of the half-light size of young star clusters give typical values in the range of a few parsecs to about $0.5 \, \textrm{pc}$ \citep[e.g.][]{bastian2012a,ryon2015a,brown2021c} but the scatter at similar cluster ages is large, going up to $\approx 20 \, \textrm{pc}$ for W{\,}3 in NGC{\,}7252 \citep[][]{maraston2004a,fellhauer2005a}.
This could indicate that the initial half-mass radius of clusters is similar and diverges due to a different initial evolutionary phase given the local environment, or that other parameter influence the initial half-mass radius, such as cluster mass or the local physical conditions.
\citet{reina-campos2023b} tested different prescriptions of initial half-mass radii, considering constant values, constant densities, a linear relationships from the data provided by \citet{brown2021c}, and a theoretical model from \citet{choksi2021b} that reads
\begin{equation}
    r_{\mathrm{c} ,\, h} = \bigg( \frac{3}{10\pi^{2}} \frac{\alpha_{\mathrm{vir}}}{\phi_{P} \phi_{\overline{P} ,\, j}} \frac{m_{\mathrm{c}}^{2}}{\Sigma_{\mathrm{g}}^{2}} \bigg)^{1/4} \frac{\epsilon_{\mathrm{c}}^{1/2}}{2\epsilon_{c} - 1} \frac{f_{\mathrm{acc}}}{Q_{\mathrm{eff}}^{2}} \;,
    \label{equ:cluster_radii_env}
\end{equation}
with $f_{\mathrm{acc}} = \num{0.6}$, $\epsilon_{c} = 1.0$, and environmentally-dependent parameters evaluated for each ring.\footnote{The parameter $\epsilon_{c}$ refers to the efficiency of star formation within a giant molecular cloud and not the average efficiency of star formation within the interstellar medium that was introduced in \Cref{subsubsec:bound_fraction_of_star_formation}.}
\citet{reina-campos2023b} conclude that none of the prescriptions reproduce the size-mass relationship after evolving the cluster population for a few Gyr within the \textit{EMP-Pathfinder} simulation suite \citep[][]{reina-campos2022b}.

Similar to \citet{reina-campos2023b} we implement different prescriptions for the initial half-mass radius, that we outline in \Cref{tab:appendix_half-mass_radius}, to test their influence on the distributions of stellar mass and half-mass radii, as discussed in \Cref{subsec:stellar_masses,subsec:half-mass_radii}.

We find that the precise details of the initial half-mass radius affects the star cluster properties (see \Cref{fig:variations_cluster_radii,fig:variations_cluster_masses} below).
When we choose a more compact initial half-mass radius, the star clusters expand more quickly at young ages, roughly matching the distribution of young star clusters in the observational data set of \citet{brown2021c} in dwarf galaxies.
However the agreement in the half-mass radii becomes worse in more massive galaxies and the slope of the $M_{\textrm{v}}$-SFR relationship steepens as less massive star clusters loose more mass, resulting in an increase of $M_{\textrm{V}}$ at low star formation rates.

When using \Cref{equ:cluster_radii_env}, we find a similar distribution compared to the fiducial model as many star clusters have initial half-mass radii only slightly smaller than $0.1 \, \textrm{pc}$ given the weak dependence on star cluster mass and the inverse squared dependence on the Toomre stability parameter.
As a consequence, similar to the fiducial model, this choice of initial half-mass radii fails to reproduce the observational constraints by \citet{brown2021c}.

Looking at the masses of the star clusters we find stronger deviations for the ``compact'' and ``theoretical'' scenario.
In the first one, star clusters start out compact and expand quickly in their dense environment, which results in an increase in mass-loss.
In turn, this effect shifts the distribution of star cluster masses towards lower values.
For the second scenario, the initial half-mass radius is proportional to the square-root of the star cluster mass.
Therefore, low-mass star clusters expand rapidly and lose mass quickly, directly affecting the mass distribution.
Note that, as the environment is denser in more massive disks, the mass distribution is most strongly affected in massive disk galaxies.
\begin{table}
    \caption{%
        Variations in the initial half-mass radii of star clusters.
    }
    \begin{center}
        \begin{threeparttable}
            \begin{tabular}{ll}
                \toprule
                \multicolumn{1}{c}{Model} & \multicolumn{1}{c}{Prescription} \\
                \midrule
                {Fiducial} & {$r_{\mathrm{c} ,\, h} = 1.0 \, \textrm{pc}$} \\ \addlinespace
                {Compact} & {$r_{\mathrm{c} ,\, h} = 0.1 \, \textrm{pc}$} \\
                {Density} & {$r_{\mathrm{c} ,\, h} = 1 \, \textrm{pc} \times (m_{\mathrm{c}} / 10^{4} \, \textrm{M}_{\odot})^{0.3}$} \\
                {Empirical\tnote{(a)}} & {$r_{\mathrm{c} ,\, h} = 2.37 \, \textrm{pc} \times (m_{\mathrm{c}} / 10^{4} \, \textrm{M}_{\odot})^{0.18}$} \\
                {Theoretical\tnote{(b)}} & {$r_{\mathrm{c} ,\, h} \propto \sqrt{m_{\mathrm{c}} / \Sigma_{\mathrm{g}}}$} \\
                \bottomrule
            \end{tabular}
            \begin{tablenotes}
                \item[(a)] Fitting results provided by \citet{reina-campos2023b} for the data of \citet{brown2021c} when limiting the star cluster population to ages $1 \, \textrm{Myr} \leq \tau_{\mathrm{c}} \leq 10 \, \textrm{Myr}$.
                \item[(b)] Adopted from \citet{choksi2021b}. To avoid extremely compact or extended clusters we introduce a lower and upper boundary to a star cluster's half-mass radius of $0.1$ and $100 \, \textrm{pc}$, respectively.
            \end{tablenotes}
        \end{threeparttable}
    \end{center}
    \label{tab:appendix_half-mass_radius}
\end{table}

\begin{figure*}
    \centering
    \includegraphics[width=0.9\textwidth]{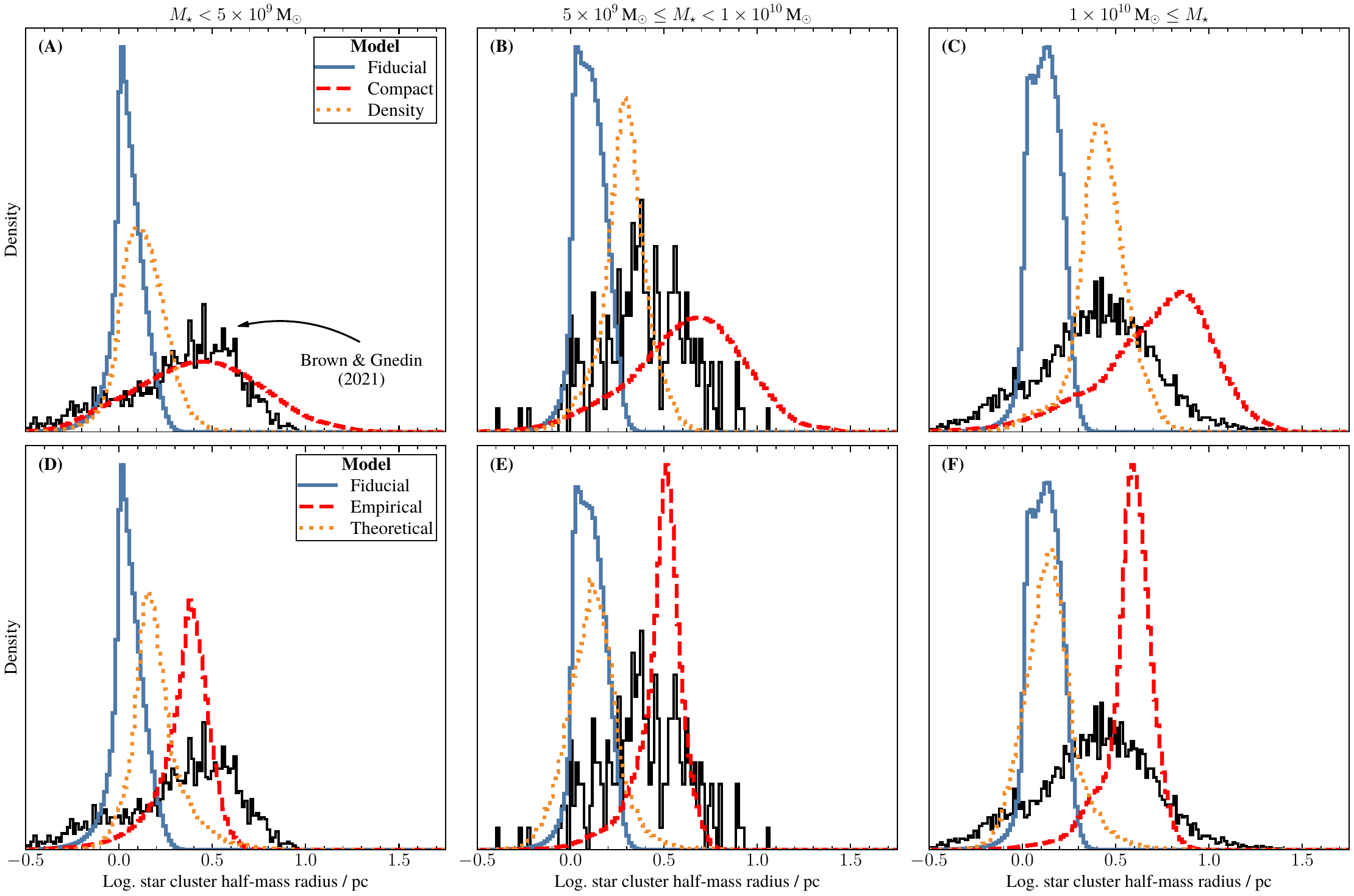}
    \caption{%
        Comparison between the distribution of half-mass radii of young star clusters in the disks of disk-dominated galaxies (see \Cref{fig:cluster_mass_distribution} for details) when considering different assumptions on the initial values, as detailed in \Cref{tab:appendix_half-mass_radius}.
        Black solid lines give the observational data set of \citet{brown2021c}.
    }
    \label{fig:variations_cluster_radii}
\end{figure*}

\begin{figure*}
    \centering
    \includegraphics[width=0.9\textwidth]{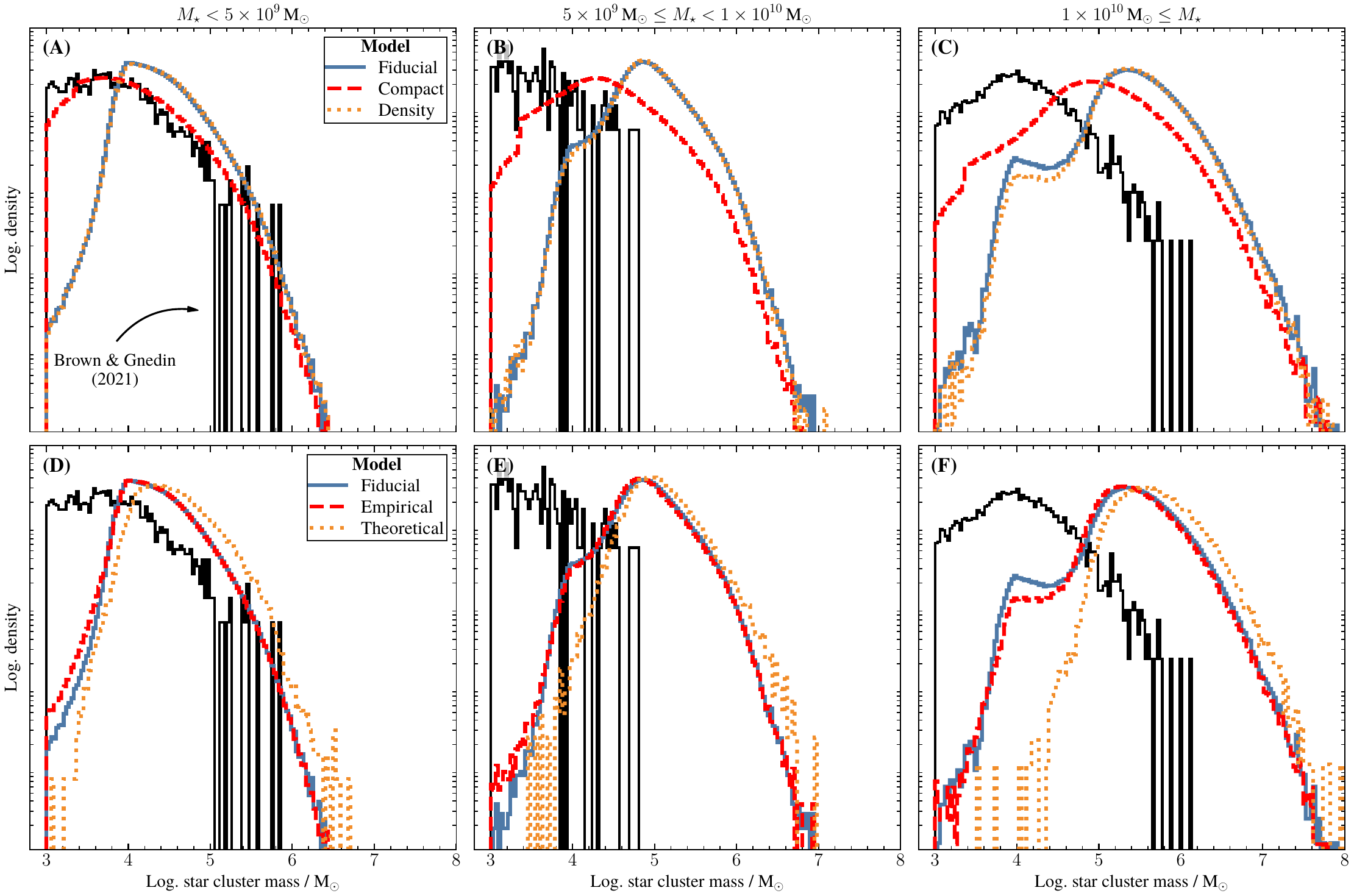}
    \caption{%
        Same as \Cref{fig:variations_cluster_radii} but for star cluster masses.
    }
    \label{fig:variations_cluster_masses}
\end{figure*}

\subsection{Cluster initial mass function}
\label{subsec:cluster_initial_mass_function}

To test the effect of the upper truncation mass on the resulting star cluster populations we re-run the model considering a pure power-law cluster initial mass function, i.e.\ $\xi(m_{\mathrm{c}}) \propto m_{\mathrm{c}}^{\alpha}$.

While the overall shape of the half-mass radius distribution remains unchanged, we find that star cluster masses old the old population are elevated by $0.2$ to $0.3 \, \textrm{dex}$.
This rather small difference of a factor of two in star cluster mass comes from the relatively high upper truncation mass of the cluster initial mass function at high-$z$.
A small difference of the same magnitude is still present in the population of young star clusters, however, the distribution has a longer tail towards higher stellar masses compared to the fiducial model.

This result translated directly into the $M_{\textrm{V}}$-SFR relationship as well where the slope value increases (from $\alpha \approx -2.47$ for the fiducial model to $\alpha \approx -2.33$ for the model variation) and absolute magnitudes are slightly increased.
As a consequence, the model variation now fails to account for some of the lowest-mass star clusters in this parameter space around star formation rates of $1 \, \textrm{M}_{\odot} \, \textrm{yr}^{-1}$.
Other parameter distributions, such as the mean star cluster metallicity versus galaxy mass remain unchanged.

\end{document}